\documentclass[preprint, times]{aastex62}

\usepackage{amsmath, graphicx, amssymb, subfigure}
\usepackage[normalem]{ulem}
\usepackage[colorinlistoftodos,textwidth=1.2cm]{todonotes}
\graphicspath{{./}{figures/}}

\received{2020 August 1}
\revised{2020 December 1}
\accepted{2020 December 2}

\submitjournal{ApJ}

\shortauthors{Michail et al.}
\shorttitle{OMC-1 Far-Infrared Polarization Spectra}

\begin{document}
\title{Far-Infrared Polarization Spectrum of the OMC-1 Star-Forming Region}
\correspondingauthor{Joseph M. Michail}
\email{michail@u.northwestern.edu}

\author[0000-0003-3503-3446]{Joseph M. Michail}
\affil{Department of Astrophysics and Planetary Science, 800 E. Lancaster Ave., Villanova University, Villanova, PA 19085, USA}
\affil{Department of Physics, Villanova University, 800 E. Lancaster Ave., Villanova, PA 19085, USA}
\affil{Center for Interdisciplinary Exploration and Research in Astrophysics (CIERA) and Department of Physics and Astronomy, Northwestern University, 1800 Sherman Ave., Evanston, IL 60201, USA}

\author{Peter C. Ashton}
\affil{Department of Physics, University of California, Berkeley, 366 LeConte Hall, Berkeley, CA 94720, USA}
\affil{Physics Division, Lawrence Berkeley National Laboratory, 1 Cyclotron Rd. Berkeley, CA 94720, USA}
\affil{Kavli Institute for the Physics and Mathematics of the Universe (WPI), Berkeley Satellite, University of California, Berkeley, Berkeley CA 94720, USA}

\author{Marc G. Berthoud}
\affil{Center for Interdisciplinary Exploration and Research in Astrophysics (CIERA) and Department of Physics and Astronomy, Northwestern University, 1800 Sherman Ave., Evanston, IL 60201, USA}

\author[0000-0003-0016-0533]{David T. Chuss}
\affil{Department of Physics, Villanova University, 800 E. Lancaster Ave., Villanova, PA 19085, USA}

\author{C. Darren Dowell}
\affil{NASA Jet Propulsion Laboratory, California Institute of Technology, 4800 Oak Grove Drive, Pasadena, CA 91109, USA}

\author[0000-0001-8819-9648]{Jordan A. Guerra}
\affil{Department of Physics, Villanova University, 800 E. Lancaster Ave., Villanova, PA 19085, USA}

\author{Doyal A. Harper}
\affil{Department of Astronomy and Astrophysics, 5640 S. Ellis Ave, Chicago, IL 60637, USA}

\author[0000-0003-1288-2656]{Giles Novak}
\affil{Center for Interdisciplinary Exploration and Research in Astrophysics (CIERA) and Department of Physics and Astronomy, Northwestern University, 1800 Sherman Ave., Evanston, IL 60201, USA}

\author{Fabio P. Santos}
\affil{Max-Planck-Institute for Astronomy, K\"onigstuhl 17, D-69117 Heidelberg, Germany}

\author{Javad Siah}
\affil{Department of Physics, Villanova University, 800 E. Lancaster Ave., Villanova, PA 19085, USA}

\author[0000-0002-1106-4881]{Ezra Sukay}
\affil{Department of Astronomy and Astrophysics, 5640 S. Ellis Ave, Chicago, IL 60637, USA}

\author{Aster Taylor}
\affil{Department of Astronomy and Astrophysics, 5640 S. Ellis Ave, Chicago, IL 60637, USA}

\author[0000-0002-6488-8227]{Le Ngoc Tram}
\affil{Stratospheric Observatory for Infrared Astronomy, Universities Space Research Association, NASA Ames Research Center, MS 232-11, Moffett Field, CA 94035, USA}
\affil{University of Science and Technology of Hanoi, Vietnam Academy of Science and Technology, 18 Hoang Quoc Viet, Vietnam}

\author[0000-0001-8916-1828]{John E. Vaillancourt}
\affil{Lincoln Laboratory, Massachusetts Institute of Technology, 244 Wood Street, Lexington, Massachusetts 02420-9108, USA}

\author[0000-0002-7567-4451]{Edward J. Wollack}
\affil{NASA Goddard Space Flight Center, Greenbelt, MD 20771, USA}

\begin{abstract}
   We analyze the wavelength dependence of the far-infrared polarization fraction toward the OMC-1 star forming region using observations from HAWC+/SOFIA at 53, 89, 154, and 214 \micron. We find that the shape of the far-infrared polarization spectrum is variable across the cloud and that there is evidence of a correlation between the slope of the polarization spectrum and the average line-of-sight temperature.  The slope of the polarization spectrum tends to be negative (falling toward longer wavelengths) in cooler regions and positive or flat in warmer regions.  This is very similar to what was discovered in $\rho$ Oph A via SOFIA polarimetry at 89 and 154 \micron. Like the authors of this earlier work, we argue that the most natural explanation for our falling spectra is line-of-sight superposition of differing grain populations, with polarized emission from the warmer regions and less-polarized emission from the cooler ones. 
   In contrast with the earlier work on $\rho$ Oph A, we do not find a clear correlation of polarization spectrum slope with column density. This suggests that falling spectra are attributable to variations in grain alignment efficiency in a heterogeneous cloud consistent with radiative torques theory.  Alternative explanations in which variations in grain alignment efficiency are caused by varying gas density rather than by varying radiation intensity are disfavored.  
\end{abstract}


\section{Introduction} \label{sec:intro}
Observing magnetic fields in molecular clouds is difficult, and many open questions surround the role that these fields play in star formation \citep{Crutcher2012,Li2014,Pattle2019}.  Measurements of polarized far-infrared (FIR) and submillimeter thermal emission from dust grains aligned with the magnetic field allow for the directions of that field to be traced within molecular clouds.  From such data, researchers have estimated the strength of a cloud's magnetic field based on diagnostics such as the degree of order in the inferred field direction \citep{Hildebrand2009,Houde2009,Houde2011,Houde2016} and the correlation between field direction and orientation of elongated structures \citep{PlanckXXXV2016,Hull2017}. For sight lines where background sources are detectable through a molecular cloud, optical and near-infrared (NIR) polarimetry can provide another method for probing aligned dust grains \citep{Sugitani2011}.  Evidence from both FIR/submillimeter polarized emission and background starlight polarized by extinction suggests that dust grains residing in dense regions are well shielded from starlight and have lower polarization efficiency. This implies that these grains are more poorly aligned or may be in an environment that causes the grain shapes to be less elongated  \citep{Arce1998,Andersson2015}. 
\citet{Hildebrand1999} showed that another technique for probing systematic changes in dust polarization efficiency is to observe the variations of polarization fraction with wavelength $p(\lambda)$ normalized to a reference wavelength $\lambda_0$, that is, $p(\lambda) / p(\lambda_0)$. Normalizing the polarization spectrum to a single wavelength eliminates sensitivity to depolarizing effects that affect all wavelengths equally. For example, the inclination of the magnetic field to the line of sight \citep{Hildebrand1999} for a given sight line would uniformly reduce the polarization over all wavelengths while preserving the shape of the spectrum. Here we present an analysis of four-band FIR polarization spectra observed in the \object{OMC-1} star forming region.  The data were obtained using SOFIA's HAWC+ polarimeter at 53, 89, 154, and 214 \micron\, and have been discussed by \citet{Chuss2019}. 

The favored explanation for magnetic alignment of dust is the radiative torques (RAT) theory \citep{Dolginov1976,Draine1997,Lazarian2007, Andersson2015}. The key ingredient is the nonvanishing net radiative torque that acts on chiral grains in the presence of an anisotropic radiation field.  Among the theory's successes is its natural explanation for the above-mentioned loss of polarization efficiency for grains shielded from starlight.  This effect arises because radiative torques preferentially operate when the aligning radiation wavelength, $\lambda$, is comparable to or less than the particle size, $a$, ($\lambda < 2a$), that is, in the UV-to-NIR range for typical dust grain sizes $<1$ \micron~ \citep[][and references therein]{Andersson2015}.   

One way to detect the loss of polarization efficiency for dense, shielded regions is to study the dependence of $p$ on column density $N$. This assumes that geometries are sufficiently simple that column density is a good tracer of dense, cold gas. For example, \citet{Arce1998} found an essentially flat curve when they plotted $p$ vs.\ selective extinction, $E(B-V)$ (a proxy for column density), for NIR polarimetry of background stars seen through the Taurus Molecular Cloud. This result implies that polarization efficiency decreases with increasing column density because a uniformly aligned grain population would have dichroic extinction proportional to $A_\text{V}$ \citep{Andersson2015}.  \citet{Arce1998} concluded that much of the interior volume of Taurus contains grains with very low polarization efficiency.  Similar conclusions have been reached from dust emission polarimetry studies that show $p$ vs.\ $N$ falling steeply \citep{Matthews2001,Fissel2016}. A careful treatment of this problem has to also consider the possibility that turbulence increases with increasing column density, which can also lead to anticorrelation between $p$ and $N$ \citep{Jones2015,PlanckXII}.

\citet{Hildebrand1999} presented FIR polarization spectra and argued that the loss of polarization efficiency for dense, cold regions could be seen in these observations.  To understand their argument, one can consider a sight line along which two different populations of grains are found: (1) warm grains with high polarization efficiency and (2) cold grains with low polarization efficiency.
Relatively speaking, the warm population will emit radiation having high polarization fraction and will be the dominant source of emission at short FIR wavelengths, while the cold population will emit radiation having small polarization fraction and will dominate the emission at long FIR wavelengths.  The result is a negatively sloped $p$ vs.\ $\lambda$ curve, as observed by \citet{Hildebrand1999}. 
Following the terminology used by \citet{Hildebrand1999}, we will refer to this line-of-sight superposition effect that drives FIR polarization spectrum slopes toward negative values as the ``heterogeneous cloud effect'' or HCE. The favored explanation for the HCE is based in RAT theory. In this scenario, grains in warmer regions are well aligned due to their exposure to the anisotropic radiation field required for RAT alignment.  Grains in dense, cool regions are shielded from this radiation and are thus poorly aligned.

The term HCE is introduced here as a more general term for the extinction-temperature-alignment correlation (ETAC) that has been studied in previous papers \citep{Ashton2018,Santos2019}. These earlier papers focused on clouds without embedded sources such that the warm, aligned grains tend to reside near the cloud surfaces. Thus, this past work equated the extinction along a given sight line, proxied by column density, to the shielding of the grains from the radiation source responsible for the alignment.  The change in nomenclature that we suggest here is motivated by an extension of the physics of ETAC to clouds with more general geometries between radiation sources and dust grains. Specific to this paper, this applies to clouds with embedded sources for which grains deep within the clouds might be expected to be aligned.  This change in nomenclature is intended to clarify the general picture that is emerging for polarization spectrum studies--that the grains in warm regions are better aligned than the cold ones, resulting in a falling polarization spectrum for sight lines where both exist. This general picture is supported by arguments we make in this paper.

It is important to note here that RAT alignment is not the only possible explanation for the HCE. In principle, it is possible that volume density, not the temperature,  is the important parameter controlling the polarization efficiency. For example, high volume density could lead to more gas-grain collisions that might decrease the alignment \citep{Andersson2015}. High density might also lead to grain size growth due to coagulation \citep{Ysard2013} that might result in rounder grains, emitting radiation of lower polarization fraction. In this paper, we will generally assume that the HCE is in fact due to the action of radiative torques, but in Section \ref{sec:discussion} we will again consider alternative explanations.

Dust in the diffuse ISM is subject to very little radiation shielding, so the corresponding FIR/submillimeter polarization spectra are generally assumed to be unaffected by HCE.  For example, \citet{Ashton2018} observed the submillimeter polarization spectrum of a translucent molecular cloud in Vela and argued that the HCE for their observations is negligible.  For the diffuse ISM, the shape of $p$ vs.\ $\lambda$ is expected to be determined only by the properties of the dust grain populations (vs. also by environment), which are assumed to be spatially homogeneous. Theoretical models for diffuse ISM polarization spectra have been presented by \citet{Draine2009}, \citet{Guillet2018}, and \citet{Lee2019}.  In contrast with the negatively sloped FIR polarization spectra observed by \citet{Hildebrand1999}, the FIR portions of these predicted diffuse ISM polarization spectra are either flat or positively sloped, except for unusual regions having very strong radiation fields, where negative slopes can be seen for the longer FIR wavelengths, such as longward of 100 \micron.  Using data from the Planck satellite and the BLASTPol balloon-borne polarimeter, observers have constructed diffuse cloud polarization spectra for comparison with models \citep{PlanckXXII,Ashton2018}.  Overall, the observed spectra are remarkably flat over the portion of the spectrum probed, which extends from 250 \micron\ to beyond 1 mm.  This flatness appears to challenge at least some of the models.

In dense molecular clouds, observations show negative slopes in the FIR \citep{Hildebrand1999,Vaillancourt2008,Zeng2013} that are consistent with the HCE and have been attributed to grain alignment efficiency being a function of radiative environment.  Longward of 250 \micron\ the situation is more complicated, with low-resolution whole-cloud maps showing flat spectra \citep{Gandilo2016,Shariff2019}, while high-resolution ground-based maps that are generally restricted to very high column densities show a transition from negative to positive slope beginning around 350 \micron\ \citep{Vaillancourt2008,Vaillancourt2012,Zeng2013}.  This transition is not well understood.  \citet{Bethell2007} presented predicted polarization spectra for dense molecular clouds, derived from magnetohydrodynamic numerical simulations.  Their models are based on RAT theory, and they do include the HCE (via integration over sight lines including a range of grain populations).  However, in contrast with the observations, they do not find negatively sloped polarization spectra in the FIR.

\citet{Santos2019} were the first observers to constrain the slope of the FIR polarization spectrum using SOFIA data.  Their target was $\rho$ Ophiuchi A, a molecular cloud core in L1688.  Using the 154-to-89 \micron~polarization ratio as a proxy for the slope of the FIR polarization spectrum, they found negative slopes toward the denser, colder regions near the core center, which they attributed to HCE, that is, superposition of warm, aligned and cool, non-aligned grains.  However, toward the warmer sight lines that correspond to low column density, they found evidence of (for the first time) positively sloped FIR polarization spectra.  They argued that these tenuous sight lines have no cold grains shielded from radiation, and thus no HCE acting to drive the polarization spectrum slope toward negative values. They found that their data were consistent with a quantitative model that included HCE in the following way. The dense central region of the cloud is assumed to be shielded from the radiation that is required for alignment in RAT theory. Thus, this region has low temperature and contributes no polarization to the signal. These authors embed this dense central region in a warmer shell that contains aligned grains. Sight lines that pass through both the shell and the dense central region have more negatively sloped polarization spectra than those passing only through the shell. The analysis of polarization spectra in Orion presented here enables us to test the conclusions of \citet{Santos2019} in the OMC-1 region in addition to extending spectral coverage to all four HAWC+ bands: 53, 89, 154, and 214 \micron.   

In Section \ref{sec:data} we discuss the data collected and selection criteria used in our analysis. In Section \ref{sec:analysis}, we present the polarization spectra of OMC-1 in the FIR on global to pixel-to-pixel scales and attempt to correlate their characteristics with environmental parameters. In Section \ref{sec:discussion}, we compare our results with those in \citet{Santos2019}. In Section \ref{sec:summary}, we summarize our findings and conclusions.
\section{Data}\label{sec:data}
        The High-resolution Airborne Wideband Camera+ \citep[HAWC+;][]{Harper2018} is the facility FIR photometer and polarimeter for the Stratospheric Observatory for Infrared Astronomy \citep[SOFIA;][]{Temi2018}. It is capable of photometric and polarimetric observations in four bands centered at wavelengths of 53, 89, 154, and 214~\micron. The polarimetric observations described here were performed using a nod-match-chop method where SOFIA's secondary mirror is rapidly chopped to remove a time-variable background intensity. Stokes $I$, $Q$, and $U$ are measured simultaneously using observations taken with four rotations of a stepped half-wave plate, and the data are reduced using the standard HAWC+ data reduction pipeline (DRP). See \citet{Santos2019} for a detailed summary of the DRP processing steps. The polarimetric observations used in this paper were originally published in \citet{Chuss2019}. For these data, we adopt resolutions of 5.5\arcsec, 8.9\arcsec, 15.3\arcsec, and 20.5\arcsec~for the 53, 89, 154, and 214 \micron~bands, respectively. These are slightly larger than the values cited in \citet{Chuss2019} since we take into account the additional polarimetry data smoothing performed in the DRP.
        
        To construct the polarization spectra, it is necessary to standardize the effective resolution and registration of the four datasets. For each set of finalized polarization data, the Stokes parameter and error maps are smoothed with a Gaussian kernel to a common resolution of 20.5\arcsec~to match the 214 \micron~resolution data, the lowest resolution data considered here. The kernel smoothing size is given by $\text{FWHM}_{\text{smooth}} = \sqrt{(20.5\arcsec)^2 - \text{FWHM}_{\lambda}^2}$, where $\text{FWHM}_{\lambda}$ is the native resolution of the map at wavelength $\lambda$. Next, the Stokes parameter and error maps are reprojected to a common pixelization, with four pixels per beamwidth, using a flux-conserving algorithm \citep{astropy:2013, astropy:2018}.  The polarization is debiased by
            \begin{equation}
                p = \sqrt{p_{\mathrm{m}}^2 - \sigma_p^2}.
            \end{equation}
        Here, $p_{\rm m}$ is the measured polarization defined as
            \begin{equation}
                p_m = \dfrac{\sqrt{Q^2 + U^2}}{I}.
            \end{equation}
        We have $\sigma_p$ as the measurement error of the polarization, and $p$ is the debiased polarization that will be used throughout this analysis \citep{Serkowski1974}. These are the final data that are used for the polarization spectra and are stored in a single file that includes maps of Stokes $I$, $Q$, and $U$, debiased polarization percent, polarization angle, polarized intensity, and the corresponding uncertainties for each of the wavelengths considered.
        
        There is a concern that for some regions in the Orion molecular cloud (OMC) the polarimetry data may suffer from reference beam contamination. This occurs when the telescope beams are chopped to reference regions of low intensity but with potentially large but unknown polarization fractions. This is especially true near OMC-1, which contains significant FIR emission far from the region of interest. To limit the amount of potential bias in our analysis, we use the systematic masks created in \citet{Chuss2019} that discard polarization data where the measured polarization angle is potentially affected by more than 10$^\circ$ from the intrinsic polarization angle due to reference beam contamination (see above). The mathematical formalism is described in \cite{Novak1997}; the assumptions and details of the cuts used here are described in detail in \cite{Chuss2019}. The angular cut chosen corresponds to the same level of uncertainty in the angle as polarization fraction measurements having a signal-to-noise ratio of at least 3.
        
        It is possible that polarization measurements at different wavelengths may be probing different regions having different magnetic fields and environments.  To guard against this possibility, we apply three data selection criteria that are based on those utilized in previous work on the polarization spectrum (references cited in Section \ref{sec:intro}): 1) We only use polarization data for which the signal-to-noise ratio for the debiased polarization fraction is greater than 3.  2) We include only sight lines for which polarization data are available for all wavelengths.  3) We remove data for points where the variation of polarization angles is greater than 15$^\circ$. This last cut restricts our analysis to regions for which changes in polarization fraction are more likely to be related to properties of the dust grains \citep{Vaillancourt2002} than to changes in the field geometry along the line of sight. 
        
\section{Analysis}\label{sec:analysis}

    \subsection{Polarization Spectra by Region}\label{ssec:pspec_region}
\begin{figure}[ht]
            \centering
            \includegraphics[width=4.5in]{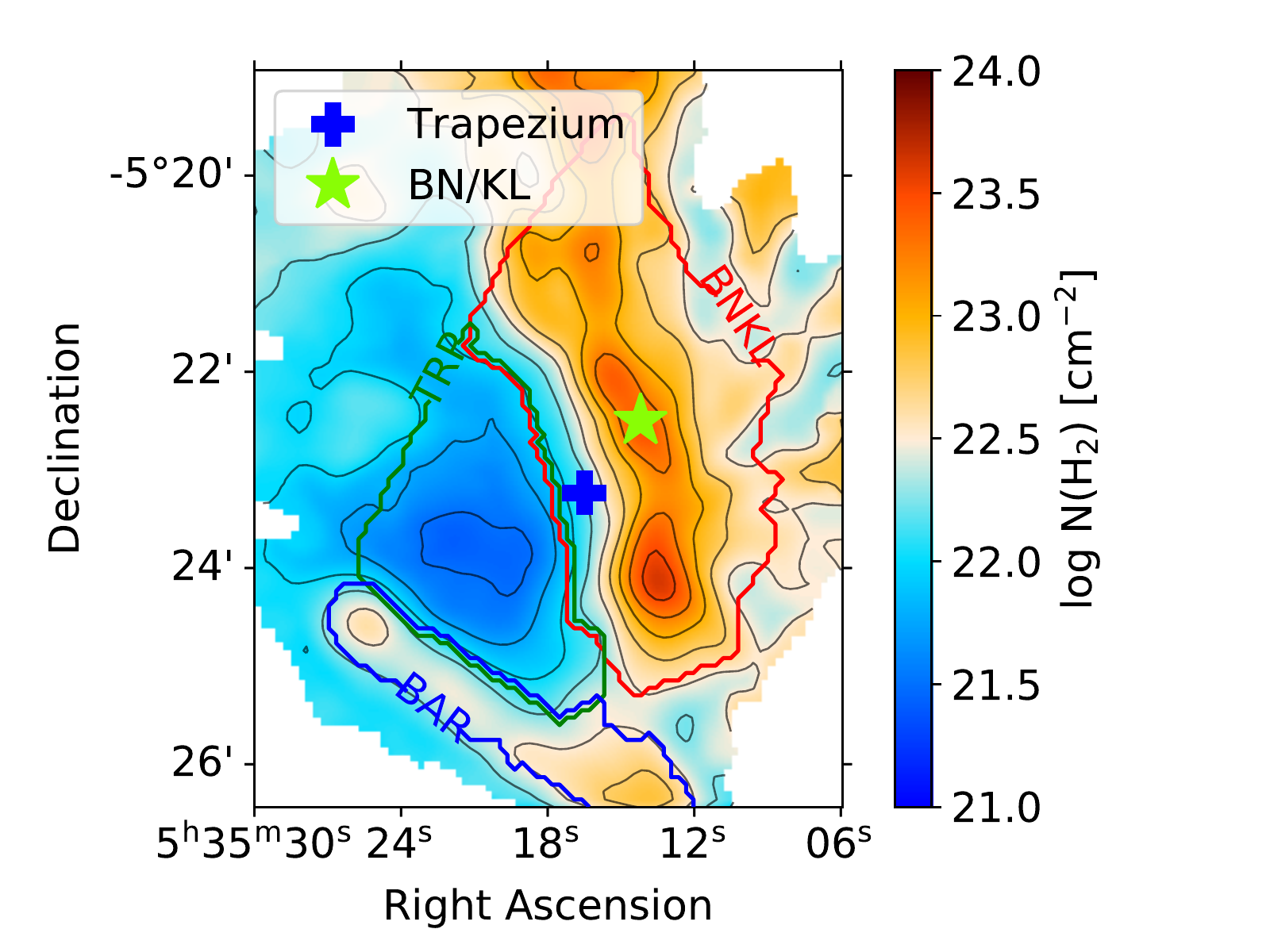}
            \caption{We consider three physically different regions corresponding to the object masks as defined by \citet{Chuss2019}. The blue cross marks the position of the Trapezium Cluster, and the green star indicates the position of the BN/KL. In color we show the logarithm of molecular hydrogen column density $N$(H$_2$) from \citet{Chuss2019}, and the contours represent 10 logarithmically spaced levels of $N$(H$_2$) between $10^{21}$ and $10^{24}$ cm$^{-2}$. The regions ``BNKL'', ``TRP'', and ``BAR'' are delineated in red, green, and blue, respectively.}
            \label{fig:masks}
        \end{figure}

     The OMC-1 region is a dynamically rich region containing wide ranges of column density, temperature, and magnetic field strength. As such, \citet{Chuss2019} split this region into three smaller regions corresponding to 1) the north-south molecular ridge (containing BN/KL and OMC-1 South), 2) the H II region formed by the Trapezium OB star association, and 3) the Orion Bar, a photodisassociation region (PDR) caused by previously mentioned Trapezium stars. The regions are referenced as ``BNKL," ``TRP," and ``BAR," respectively, and we follow this naming convention here. After data cuts, there are 2155 polarization pseudovectors in this data set, which we use to discuss and compare the differences in dust grain physics throughout these three regions. Figure \ref{fig:masks} shows these regions superposed on a map of molecular hydrogen column density $N$(H$_2$) taken from \citet{Chuss2019}. This figure shows 10 contours for the column density that are logarithmically spaced between $10^{21}$ and $10^{24}$ cm$^{-2}$. These same contours are used throughout the paper to provide a convenient spatial reference.
        
    In addition to the data cuts related to the polarization spectrum described in Section~\ref{sec:data}, for the analysis in this section, we require there be at least three pixels within a region that pass our data selection criteria in order to report a spectrum for that region. Both the BNKL and TRP regions pass this criterion; however, the BAR does not. Thus in this section we show results for the TRP and BNKL regions only. We do not exclude BAR pixels from cloud-wide analyses, however. The lack of consistency of polarization direction with wavelength over most of the BAR region could be due to a combination of high turbulence, which results in low polarization fractions, and varying reference beam contamination over the four HAWC+ bands \citep{Chuss2019}. 
        
\begin{figure*}
        \centering
        \subfigure[Overall Cloud Spectrum]{
            \includegraphics[width=2.2in]{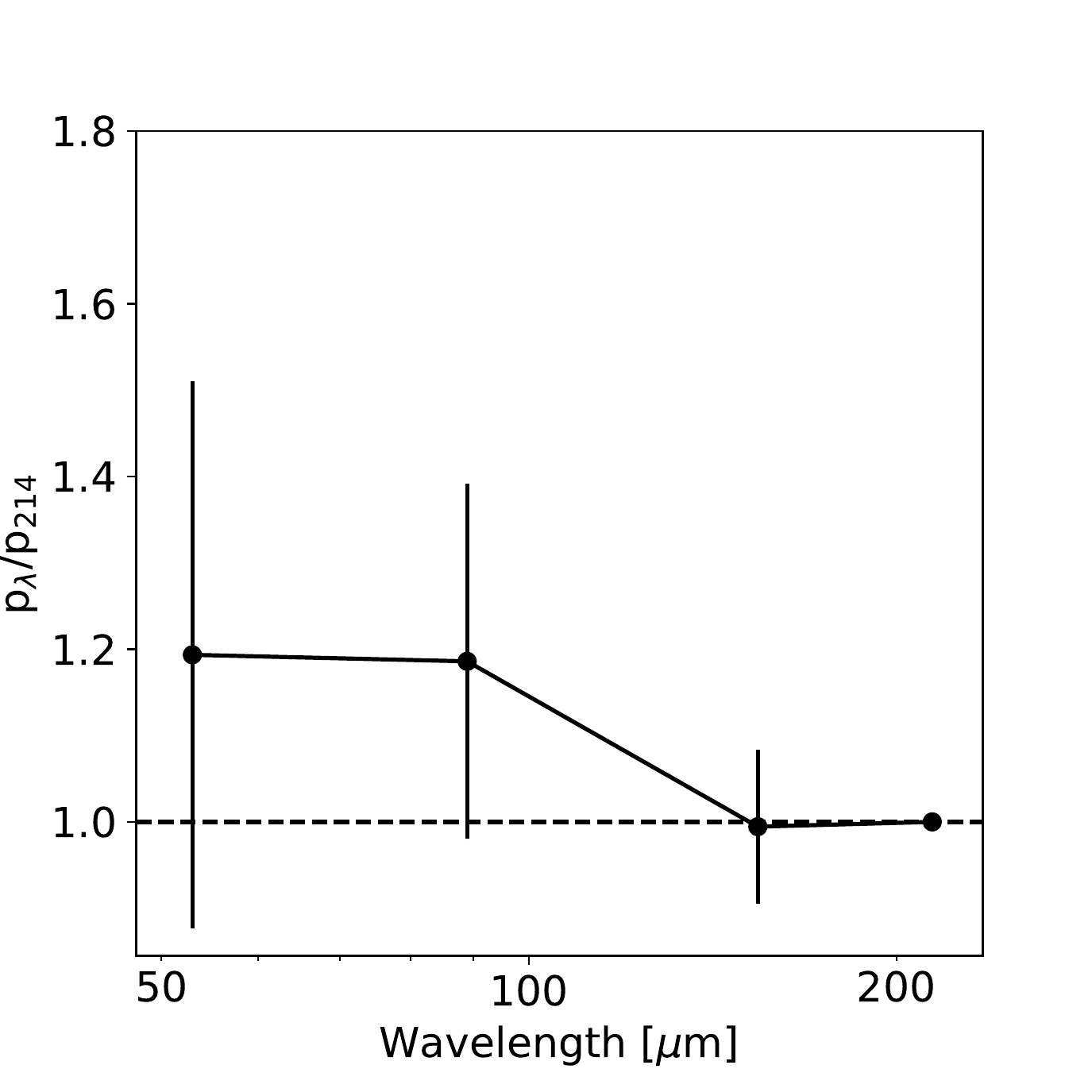}
            \label{fig:hawc_pspec_overall}
        }
        \subfigure[BNKL Region Spectrum]{
            \includegraphics[width=2.2in]{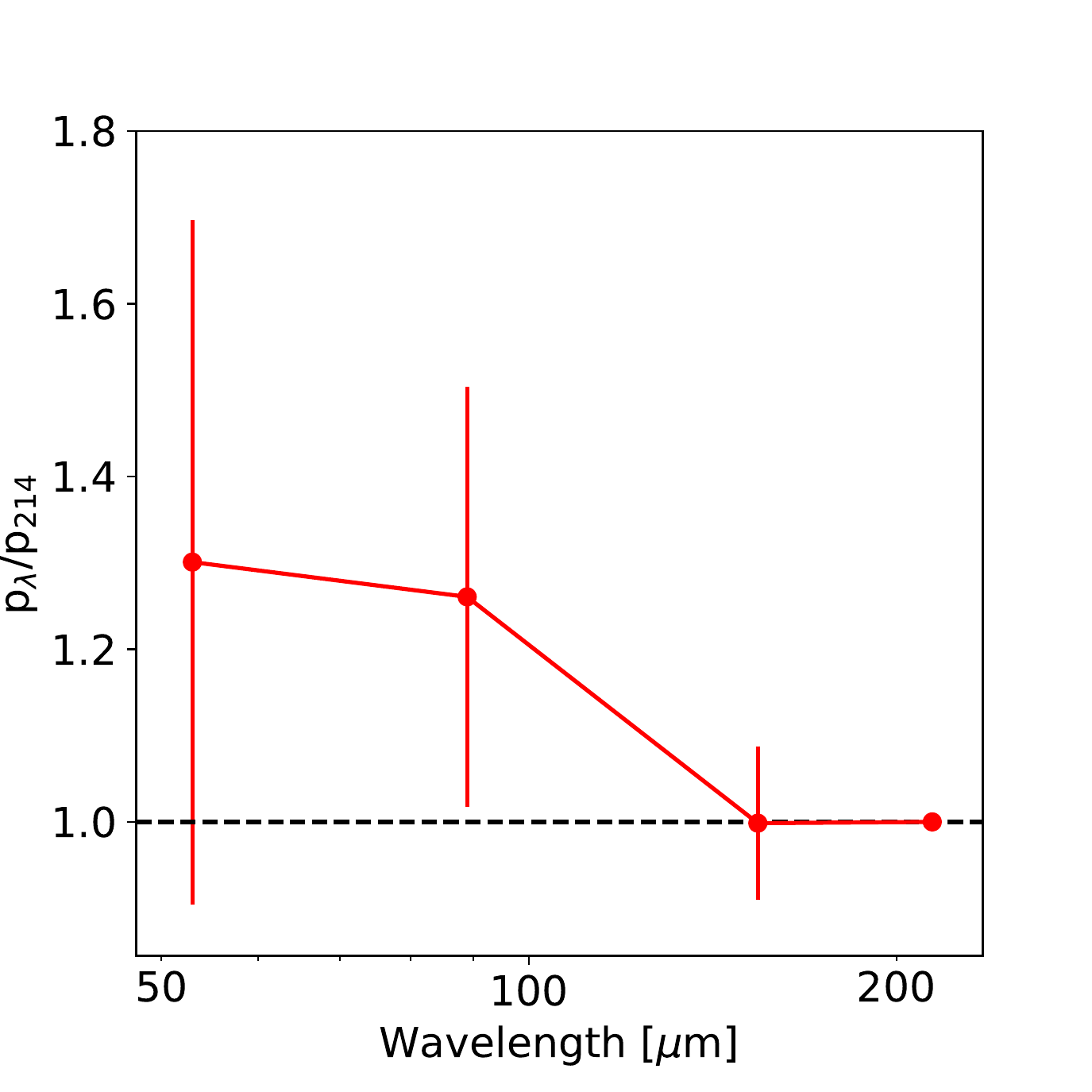}
            \label{fig:hawc_pspec_bnkl}
        }
        \subfigure[Trapezium Region Spectrum]{
            \includegraphics[width=2.2in]{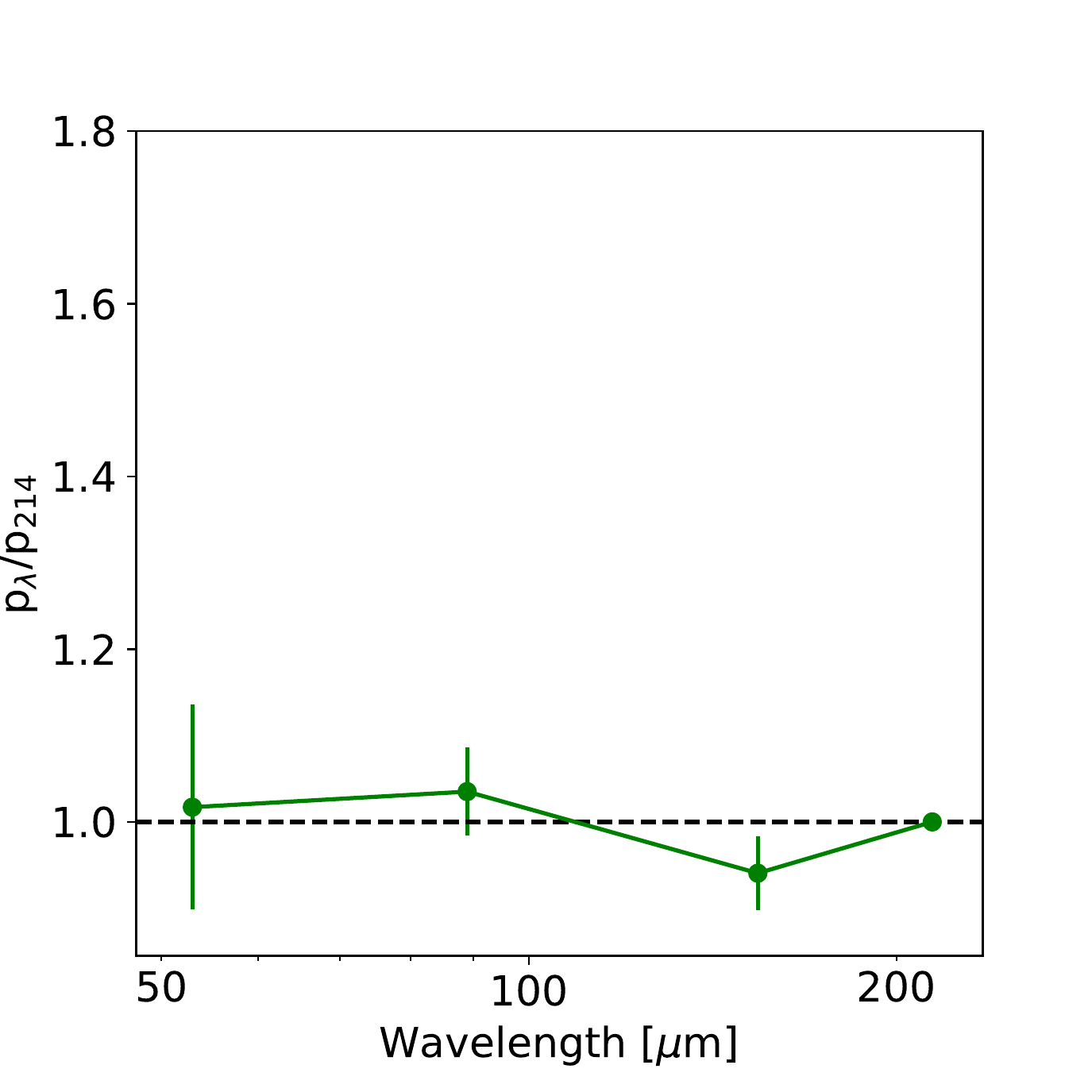}
            \label{fig:hawc_pspec_trp}
        }
        \caption{Polarization spectra for OMC-1, normalized to the 214 \micron\ polarization value, for (a) the entire cloud, (b) BNKL region, and (c) TRP region. Data are limited to sight lines for which the range of polarization angles over all four wavelengths is smaller than 15$^\circ$. The ratios presented are determined by taking the median of the normalized polarization ratios within each of our three regions. The vertical bars indicate the median absolute deviation (MAD, over the cloud/region) of the normalized ratios.}
        \label{fig:polspec_hawc}
    \end{figure*}

    Shown in Figure \ref{fig:polspec_hawc} are global polarization spectra for the entire cloud and for the BNKL and TRP regions. We follow the standard practice employed in previous polarization spectrum analyses by normalizing polarization fractions across all of our data to a common wavelength (see Section \ref{sec:intro} for more details). We normalize to the 214 \micron~wavelength as it is closest to the normalization wavelength of 350 \micron\ used in early work \citep{Vaillancourt2002}.
        
    The values reported in Table \ref{tab:pol_ratios} and Figure \ref{fig:polspec_hawc} are calculated by finding the median value of each wavelength's normalized polarization fractions within each of the defined regions. The vertical  bars indicate the median absolute deviation (MAD) of the normalized data across each region.

        \begin{deluxetable*}{cccc}
            \tabletypesize{\normalsize}
            \tablehead{\colhead{Region} & \colhead{$p_{53} / p_{214}$} & \colhead{$p_{89} / p_{214}$} & \colhead{$p_{154} / p_{214}$}}
            \startdata
                Overall  & $1.19 \pm 0.32$    & $1.19 \pm 0.21$   & $0.99 \pm 0.09$  \\
                BNKL     & $1.30 \pm 0.40$    & $1.26 \pm 0.24$   & $1.00 \pm 0.09$  \\
                Trapezium& $1.02 \pm 0.12$    & $1.04 \pm 0.05$   & $0.94 \pm 0.04$  \\
            \enddata
            \caption{Polarization ratios for the three regions considered. The values reported for each region are the median polarization ratio; ranges are estimated by the median absolute deviation (MAD). }
            \label{tab:pol_ratios}
        \end{deluxetable*}
    
   Figure \ref{fig:hawc_pspec_overall} shows a large spread in the polarization spectrum for the entire cloud.  The same is true of the BNKL region (Figure ~\ref{fig:hawc_pspec_bnkl}). In both cases, the spread is likely due to physical conditions that vary across the BNKL region.  This complex region (defined in Figure~\ref{fig:masks}) contains the BN/KL object and its environs, including an explosion likely induced by the decay of stellar orbits within the nebula \citep{Bally2017}. In addition, the region also encompasses OMC-1 South, which is thought to be spatially separated from the BN/KL object along the line of sight \citep{ODell2020} and has a significantly lower temperature \citep{Chuss2019}. There is some observed tendency toward a negative polarization spectrum slope, consistent with what has been observed by earlier FIR polarization spectrum observers. This is likely due to HCE, a superposition of warmer, well-aligned grains and cooler, poorly aligned grains along the line of sight, which was proposed by \citet{Hildebrand1999} and modeled quantitatively by \citet{Santos2019}. 

    The polarization spectrum of the TRP region shows less variation than that of the BNKL region, likely due to its relative physical uniformity. The polarized emission is likely coming from dust associated with the PDR behind the H II region \citep{ODell2020}. The column density is low here with a value of $N$(H$_2)\sim 4~\times~10^{21}$ cm$^{-2}$, corresponding to $A_\text{V}\sim4$ mag, where we have used the conversion $A_\text{V}$/$N_\text{H} = 5.3~\times~10^{-22}$ mag cm$^{2}$ \citep{Draine2011}. These low levels of dust extinction are comparable to what is found in the translucent cloud in Vela studied by \citet{Ashton2018} where they find $A_\text{V} \sim 2.6$ mag.  \citet{Ashton2018} observed a flat polarization spectrum from 250 to 850 \micron\ in this cloud, by combining polarization maps from BLASTPol with Planck. 
    
    As noted in Section \ref{sec:intro}, \citet{Ashton2018} argued that due to the relatively low dust extinction levels in their cloud, the HCE mechanism should have a negligible effect on the polarization spectrum. This is because HCE relies on line-of-sight superposition of dust grain populations having differing temperatures and alignment efficiencies, due to their differing degrees of exposure to radiation, whereas the target of \citet{Ashton2018} lacks sufficient column density to provide for shielding or cooling of dust.  \citet{Ashton2018} further argued that for this reason their observations could be meaningfully compared with dust models developed for the diffuse ISM, which do not include HCE.  They compared their observations with models by \citet{Draine2009} and \citet{Guillet2018}, finding that a subset of these models, specifically those containing aligned carbonaceous grains, are generally consistent with a relatively flat polarization spectrum in the submillimeter.  However, the models predict a falloff moving toward shorter wavelengths, so they cannot explain the flat FIR polarization spectra we observe in the TRP region.
    
    A more recent set of models for diffuse ISM polarization spectra, by \citet{Lee2019}, shows a greater degree of variability with respect to the FIR slope.  In comparison with the models of \citet{Draine2009} or \citet{Guillet2018}, these more recent models extend to more intense radiation fields.  They also include disruptive RAT (RATD) processes that become important for these more intense radiation fields, as centrifugal forces imparted to the grains by photons cause the destruction of the grains.  While no single one of the many model spectra shown by \cite{Lee2019} is as flat as what we have found for the TRP region, it is not hard to imagine that a suitable superposition of several different models might approximate the observations. Note that \citet{Tram2020} has shown that RATD models can explain the variation of the polarization fraction as a function of temperature for $\rho$ Oph A, using the same SOFIA data as were studied by \citet{Santos2019}. More in-depth comparisons of our TRP polarization spectrum with the predictions of new theoretical advances such as the RATD model and ``Astrodust'' \citep{Draine2020} are warranted but beyond the scope of the present paper.

    \subsection{Pixel-by-pixel Polarization Spectra}\label{ssec:bypixel}
\begin{figure}[htbp]
            \centering
            \includegraphics[width=3.5in]{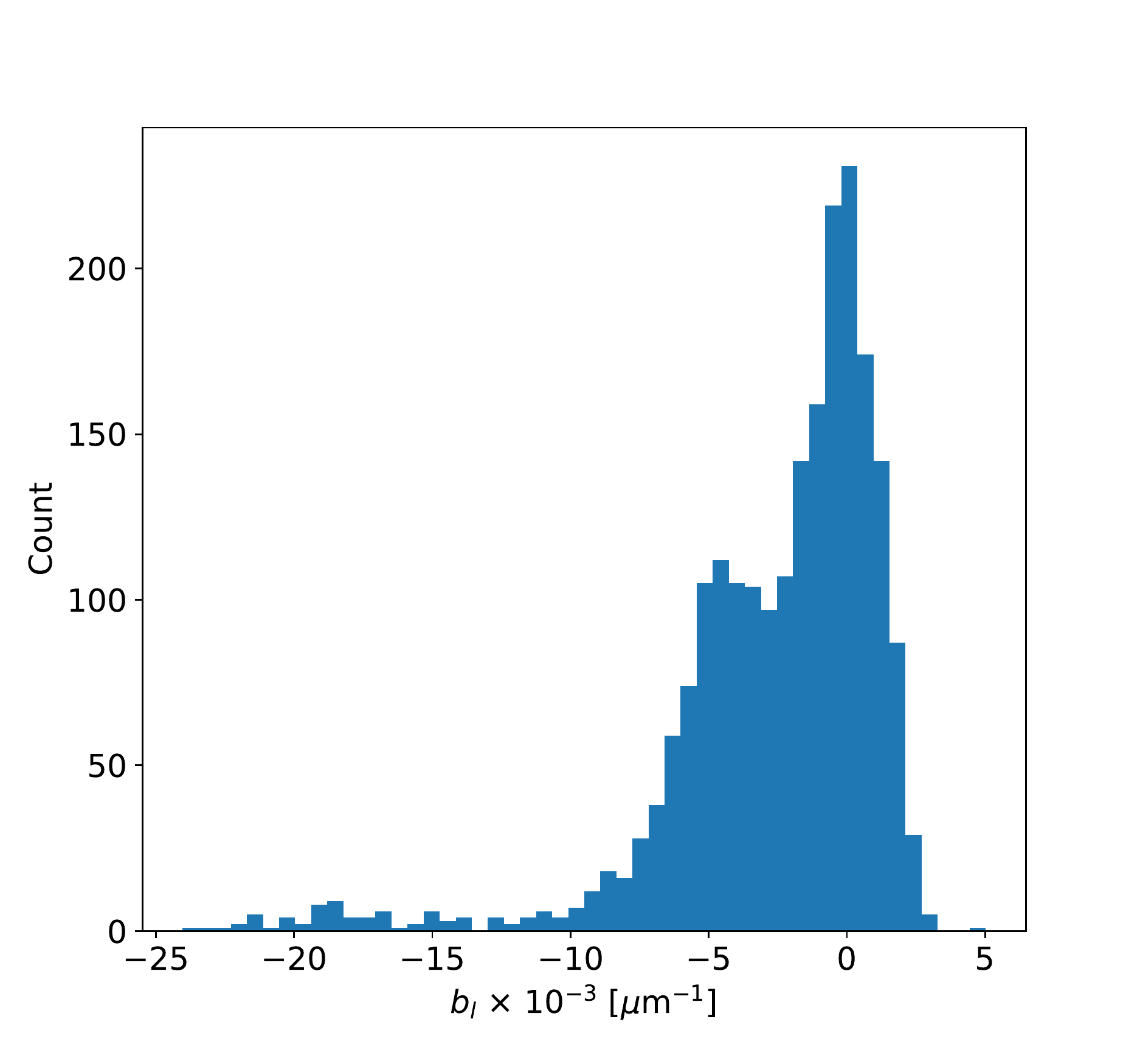}
            \includegraphics[width=3.5in]{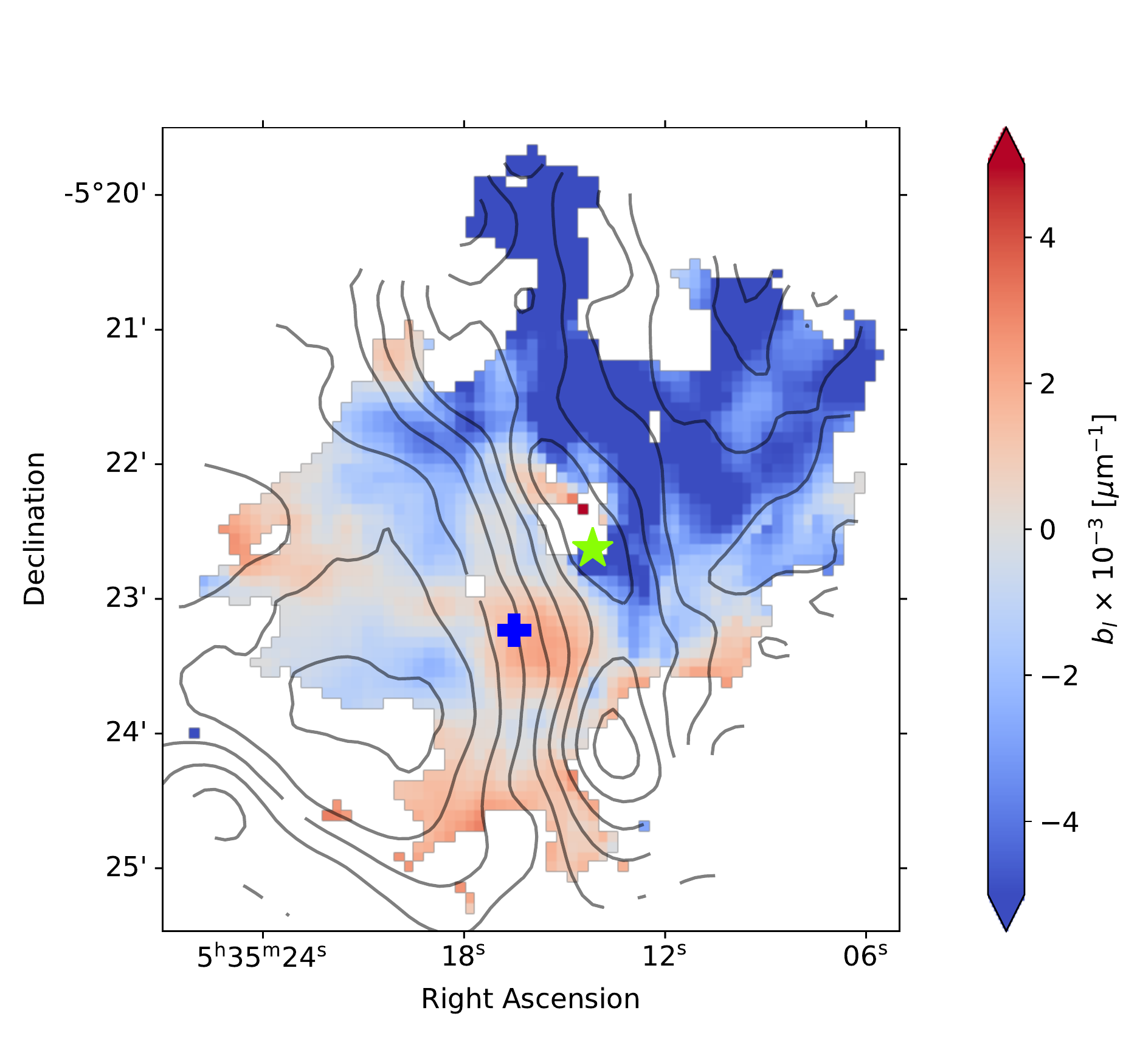}
            \caption{Left: histogram of the polarization spectra parameter $b_l$ from Equation \ref{eq:linear} for the entire cloud. This histogram employs 50 bins across the distribution. Right: spatial map of $b_l$. Column density contours in this region are shown for reference; contours are limited to $N$(H$_2$) = $10^{21 - 24}$ cm$^{-2}$ in 10 logarithmically spaced intervals. The blue cross indicates the center of the Trapezium Cluster, and the green star indicates the position of BN/KL. We saturate the color scale for values $|b_l| \geq 5\times10^{-3}$ $\mu$m$^{-1}$ to show better contrast in the spatial variability of this parameter.}
            \label{fig:fit_params}
        \end{figure}

    In view of the presumed large variability in the slope of the polarization spectrum within the BNKL region, we next turn to the characterization of polarization spectra on a pixel-by-pixel basis. In this section, we follow \citet{Gandilo2016} and \citet{Shariff2019} by regressing a model to our pixel-by-pixel polarization spectra. In this case, the choice of a linear model allows for a comparison to the analysis of \citet{Santos2019}. Again, only lines of sight that conform to the criteria listed in Section~\ref{sec:data} are considered. 
    The linear equation is described by the form
        \begin{equation}
            p(\lambda) / p(\lambda_0) = a_l\left(b_l[\lambda - \lambda_0] + 1\right).
            \label{eq:linear}
        \end{equation}
    Here, $\lambda_0$ is the normalizing wavelength that we take to be 214 \micron\ to be consistent with the analysis in Section~\ref{ssec:pspec_region}. The physically significant fit parameter here is $b_l$, which tracks the slope of the polarization spectrum\footnote{The true slope of the polarization spectrum is $a_l\cdot b_l$. However, since $a_l \sim 1$, $b_l$ is treated as the slope throughout this paper.}. A negative (positive) value of $b_l$ indicates that the polarization falls (rises) with increasing wavelength; this is referred to as a ``falling''  (``rising") spectrum. An error-weighted nonlinear least-squares regression was used to fit Equation \ref{eq:linear} to the normalized data. The averaged residual (over all wavelengths and sight lines) from our linear fit of $p(\lambda)/p(\lambda_0)$ is $0.098$, indicating well-fit spectra. We list the median and MAD values for the linear fit parameters, $a_l$ and $b_l$, in Table \ref{tab:sed_param_fit_ratios} for  the entire cloud and for the BNKL and TRP regions separately. We also plot the histogram of values for $b_l$ for the entire cloud in Figure \ref{fig:fit_params} (left). The spatial distribution of $b_l$ is shown in Figure \ref{fig:fit_params} (right). 
    
        \begin{deluxetable*}{c|cc}
            \tabletypesize{\normalsize}
            \tablehead{\colhead{Region} & \colhead{$a_l$} & \colhead{$b_l(\cdot~10^{-3})$} \\
                       \colhead{}       & \colhead{(--)}    & \colhead{($\mu$m$^{-1}$)}      }
            \startdata
                Overall  &     $0.95 \pm 0.05$  & $-1.47 \pm 2.04$\\
                BNKL     &     $0.94 \pm 0.06$  & $-2.26 \pm 2.61$\\
                TRP&     $0.98 \pm 0.03$  & $-0.36 \pm 0.74$\\
            \enddata
            \caption{Median $\pm$ MAD for linear parameter values resulting from fits to individual pixel polarization spectra within specific regions.}
            \label{tab:sed_param_fit_ratios}
        \end{deluxetable*}

    \subsection{Dependence of Polarization Spectra on Environment}\label{ssec:env}
        Insight into the physics of grain alignment can be developed by examining the polarization spectrum as a function of physical properties, namely the line-of-sight temperature and column density, at each position.
        Figure \ref{fig:fit_params} (right) shows the spatial distribution of $b_l$, obtained via the linear fits of Section~\ref{ssec:bypixel}. We have overlaid column density contours as well as marked the locations of the BN/KL and the Trapezium Cluster to more easily guide the eye. The contours again correspond to 10 logarithmically spaced bins in the range $N$(H$_2$) $=10^{21-24}$ cm$^{-2}$. The spectral parameter $b_l$ clearly has spatial coherence. To the west of the north-south molecular ridge, where the column density is relatively high, $b_l<0$ over large regions. To the east of the ridge, the sign of this parameter varies. This is consistent with the result reported in Section \ref{ssec:pspec_region}: in the eastern part of OMC-1, inside the TRP region, the median spectrum is flat with relatively little variation. 
        
        There is a nearly circular region of $b_l>0$ colocated with, but slightly offset from, the Trapezium Cluster (marked by a blue ``+''). This region approximately corresponds to the ``highly ionized region'' as designated by \citet{ODell2020}. A rising polarization spectrum here may indicate the presence of transiently heated, unaligned, small dust grains in the vicinity of the OB stars.  At the short-wavelength end of the spectrum, these grains contribute unpolarized intensity that is superposed on the partially polarized emission from the larger, aligned grains, which are presumably located in the PDR behind the H II region \citep{ODell2020}.  At the long-wavelength end of the spectrum, emission is dominated by the large dust grains that are well aligned. This is consistent with the physical assumptions underlying the models of \citet{Draine2009}.   
        
\begin{figure}[htpb]
            \centering
            \includegraphics[trim={0 1cm 0 2cm}, clip, width=5in]{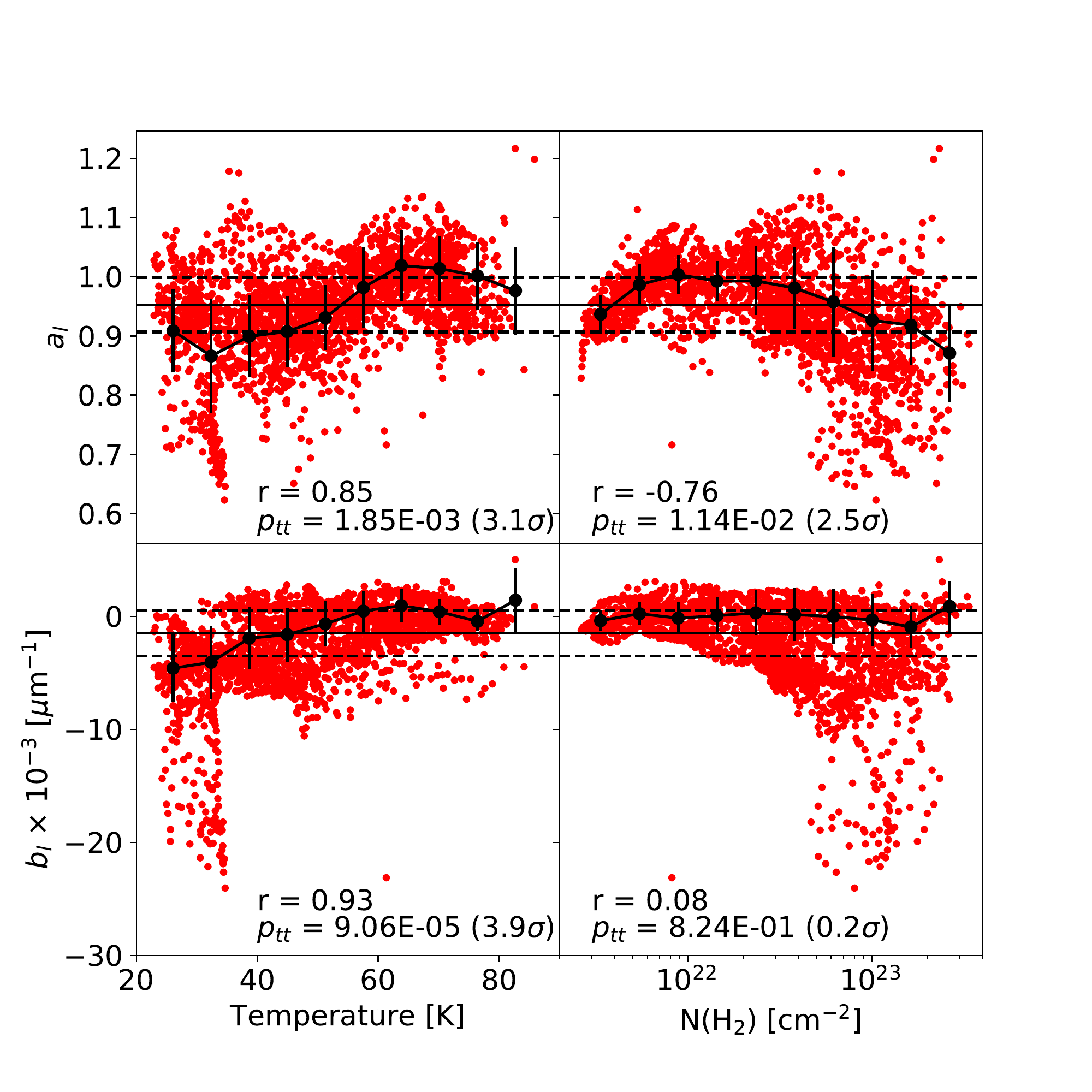}
            \caption{Correlations of linear fit parameters with SED parameters. Solid black lines are the median values of the parameters across the entire cloud. The dashed lines indicate the median $\pm$ MAD. The data are binned using a weighted mean, and the error bars indicate the weighted standard deviation within the bin. Temperature is binned linearly; column density is binned logarithmically. Here, $r$ is the error-weighted Pearson coefficient, and $p_{tt}$ is the two-tailed p-value. The significance level corresponding to $p_{tt}$ is shown in parentheses for each plot.}
            \label{fig:sedparam_linear}
        \end{figure}

        To further explore how our FIR polarization spectra depend on environment, we follow \citet{Gandilo2016} and \citet{Shariff2019} in exploring correlations between polarization spectra characteristics (in our case, the slope $b_l$) and spectral energy distribution (SED) fit parameters. For this analysis, we use the temperature and column density maps produced by \citet{Chuss2019}. These were reprojected to the grid of the 214 \micron~data that was used for the polarization spectrum work and then smoothed to 20.5\arcsec~resolution. We next bin the temperature and column density into 10 equally spaced bins (that are linear in temperature and logarithmic in column density). For each bin, we calculate the error-weighted mean and associated uncertainty of $b_l$, where we weight using the variance in the individual values of $b_l~(w = 1/\sigma^2_{b_l}$). The variance on each value of $b_l$ is estimated from the covariance matrix returned for each pixel from the fitting procedure in Section \ref{ssec:bypixel}. These results are shown in Figure \ref{fig:sedparam_linear} (lower panels). We note that low values of $a_l$ and $b_l$ correspond to sight lines where the linear model does not accurately represent the polarization spectrum. Additional terms in Equation \ref{eq:linear} are likely needed, however, which is beyond the scope of this paper.
        
        To determine the significance of the correlations, we use an error-weighted correlation test to determine the Pearson coefficient ($r$) and two-tailed p-value (denoted $p_{tt}$ to distinguish it from the polarization fraction $p$) for the binned data. The null hypothesis is the case of no correlation ({\it i.e.}, $r=0$), while the alternative hypothesis is the case of a correlation. Values of $r$ and $p_{tt}$ are noted in Figure~\ref{fig:sedparam_linear}.
        
        The bottom left panel of Figure \ref{fig:sedparam_linear} reveals a significant positive correlation between $b_l$ and temperature at a level greater than $3\sigma$. In contrast, the submillimeter polarization spectrum studies in Vela C \citep{Gandilo2016} and the Carina Nebula \citep{Shariff2019} found no correlations between polarization spectrum slope and SED parameters. We note that their observations had lower spatial resolution and longer wavelengths. As a check, we also test for correlations between the true slope ($a_l\cdot b_l$) and the SED parameters. These results (not shown) reveal that the true slope is correlated with temperature but not column density, consistent with what was found for $b_l$ as reported in Figure \ref{fig:sedparam_linear}.
        
        To determine if the correlation is robust, we complete two additional tests. The first tests the strength of the $b_l$ and $T$ correlation on the large scatter of $b_l$ values at $T \leq 35$ K. We remove all sight lines where $T \leq 35$ K and redo the fits. We find that $r=0.89,~p_{tt} = 6.50\times10^{-4}~(3.4\sigma)$, still indicating a statistical detection. Next, we bin the data using the median and take the error bar to be the MAD. This test checks the robustness of our fits to the presence of outliers. We find $r = 0.89, p_{tt} = 5.23\times10^{-4}~(3.5\sigma)$, again showing a significant correlation. Since these additional tests do not significantly affect the statistical significance of the correlation, we claim that this is a true variation of the polarization slope with temperature. Throughout these two additional tests, there was no significant correlation between $b_l$ and $N$(H$_2$).
        
        For many sight lines within Orion, the optical depth at 53 \micron, $\tau_{53}$, can be large. At 53~\micron, about 25\% of the sight lines used in this analysis have $\tau_{53} \geq 0.5$, 90\% of which lie in the BNKL region. According to \citet{Novak1989}, such large optical depths could reduce the ratio of measured-to-intrinsic polarization by more than 30\%. To check if our results are affected by this, we repeat the analysis described in Section \ref{ssec:env} while removing the sight lines where $\tau_{53} \geq 0.5$. We find that there is no major effect on the results as presented here.
        
        The work that is most directly comparable to our own is that by \citet{Santos2019}, who studied $\rho$ Oph A with a spatial resolution similar to that of this work in OMC-1. As discussed in Section~\ref{sec:intro}, they showed HCE is not observed for regions having low column density and high temperature. More specifically, they showed a positive correlation between the slope of the polarization spectrum and temperature and a negative correlation between the slope and column density. We also observe a lack of HCE for regions of low column density and high temperature. However, though we find a positive correlation between the slope of the polarization spectrum and the temperature, we observe no significant correlation between slope and column density. Recall that \citet{Santos2019} considered measurements at only two wavelengths, while we obtained four. In Section \ref{ssec:ratio_maps} below, we restrict our data to just two bands so that we may make a more direct comparison.

        \begin{figure}[htbp]
            \centering
            \includegraphics[trim=0 0.4cm 0 0.0cm, clip, width=3.5in]{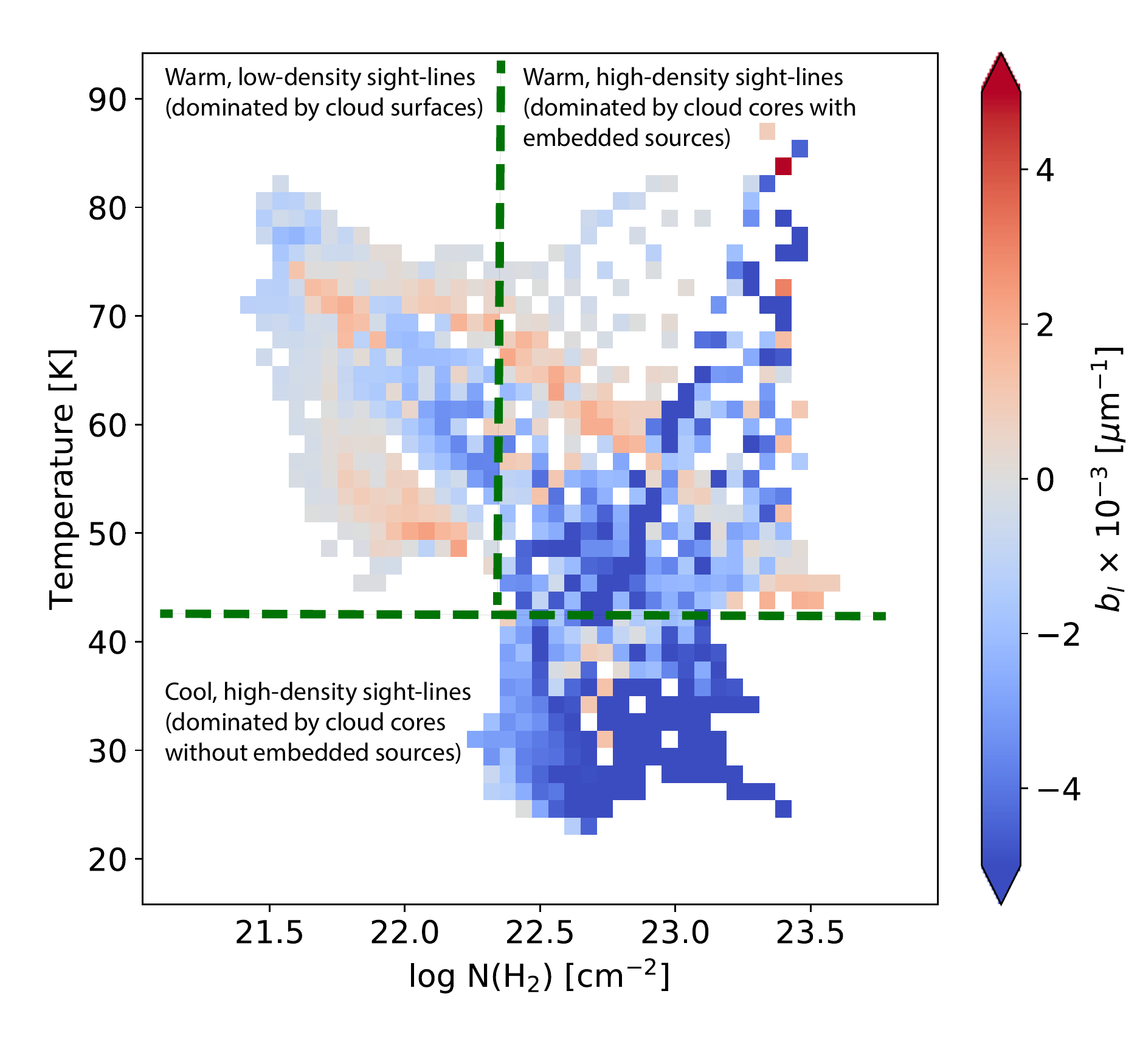}
            \includegraphics[trim=0 0 0.73cm 3.1cm, clip, width=3.5in]{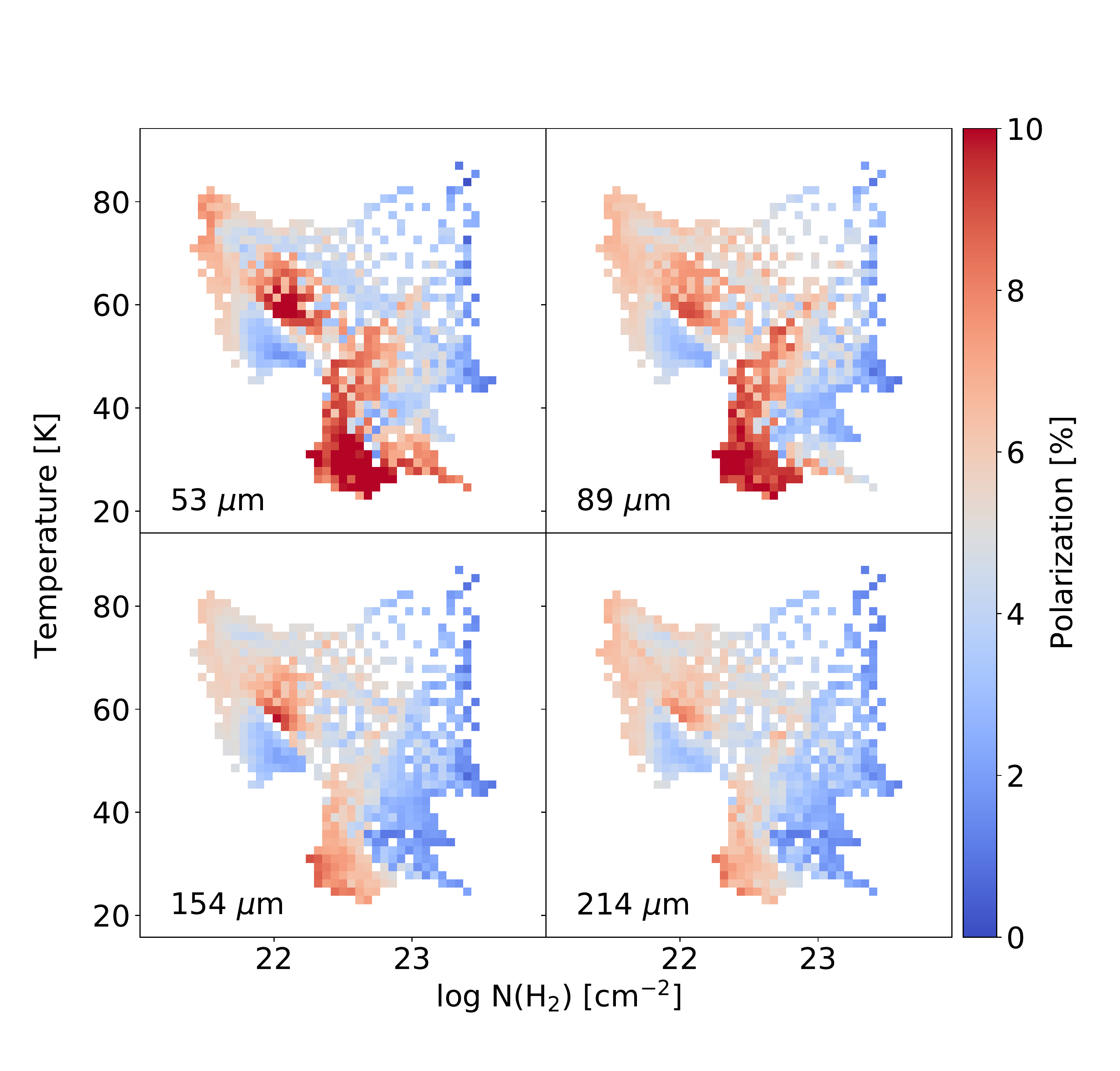}
            \caption{Left: dependence of $b_l$ on both dust temperature and column density. The temperature and logarithm of column density are binned into 50 separate bins. The color of each bin is representative of the median within the bin. Right: dependence of the polarization percent on both dust temperature and column density in each of the HAWC+ bands. The color of each bin represents the median polarization within it. }
            \label{fig:N_T_P}
        \end{figure}
        
      It is important to note that the SED fits of \citet{Chuss2019} that we use here differ in some important ways from the fitting methods used in the earlier work cited above. \citet{Gandilo2016} and \citet{Santos2019} assume a constant dust emissivity index ($\beta$) throughout their respective analyses. \citet{Gandilo2016} assumed $\beta=2$ and \citet{Santos2019} assumed a constant $\beta=1.62$ from \citet{PlanckXI}. \citet{Shariff2019} used the Planck all-sky thermal dust model where $\beta$ was treated as a free parameter. In all three cases, however, their models assumed optically thin emission. In contrast, \citet{Chuss2019} solve for temperature, column density, and $\beta$ simultaneously, making no assumptions regarding optical depth. (Exploring correlations between $b_l$ and $\beta$ is beyond the scope of this work.) Due to these fitting differences, some caution must be used when discussing the dependence of the FIR polarization spectrum slope on SED fit parameters, especially when comparing our results with those of \citet{Santos2019}.
        
      Additional insight can be gained by plotting $b_l$ as a function of both temperature and column density, as shown in Figure \ref{fig:N_T_P} (left). In this figure, the points have been grouped into three regions. The first, at the bottom of the diagram, corresponds to sight lines that are dominated by cores of clouds without significant embedded sources.  In these regions, $b_l<0$ indicating a falling spectrum, which is due to the unaligned, cool dust grains in these dominant regions.  In the upper left part of the diagram, sight lines that are dominated by cloud surfaces are represented. In these regions, the density is low and the temperature is high, and no HCE is detected.  This leads to flatter spectra than for the denser, cooler sight lines.  Finally, the upper right regions show sight lines that are dominated by dense cores for which embedded sources are present to align the grains in the denser region.  Similar to the regions dominated by cloud surfaces, the spectra are found to be on average flatter than for the cool, dense regions. These types of sight lines are likely lacking in $\rho$ Oph A \citep{Santos2019}. For completeness, we also show the polarization fraction as a function of both temperature and column density for all four wavelengths (Figure~\ref{fig:N_T_P}, right).
       
\begin{figure}[htbp]
            \centering
            \includegraphics[width=3.5in]{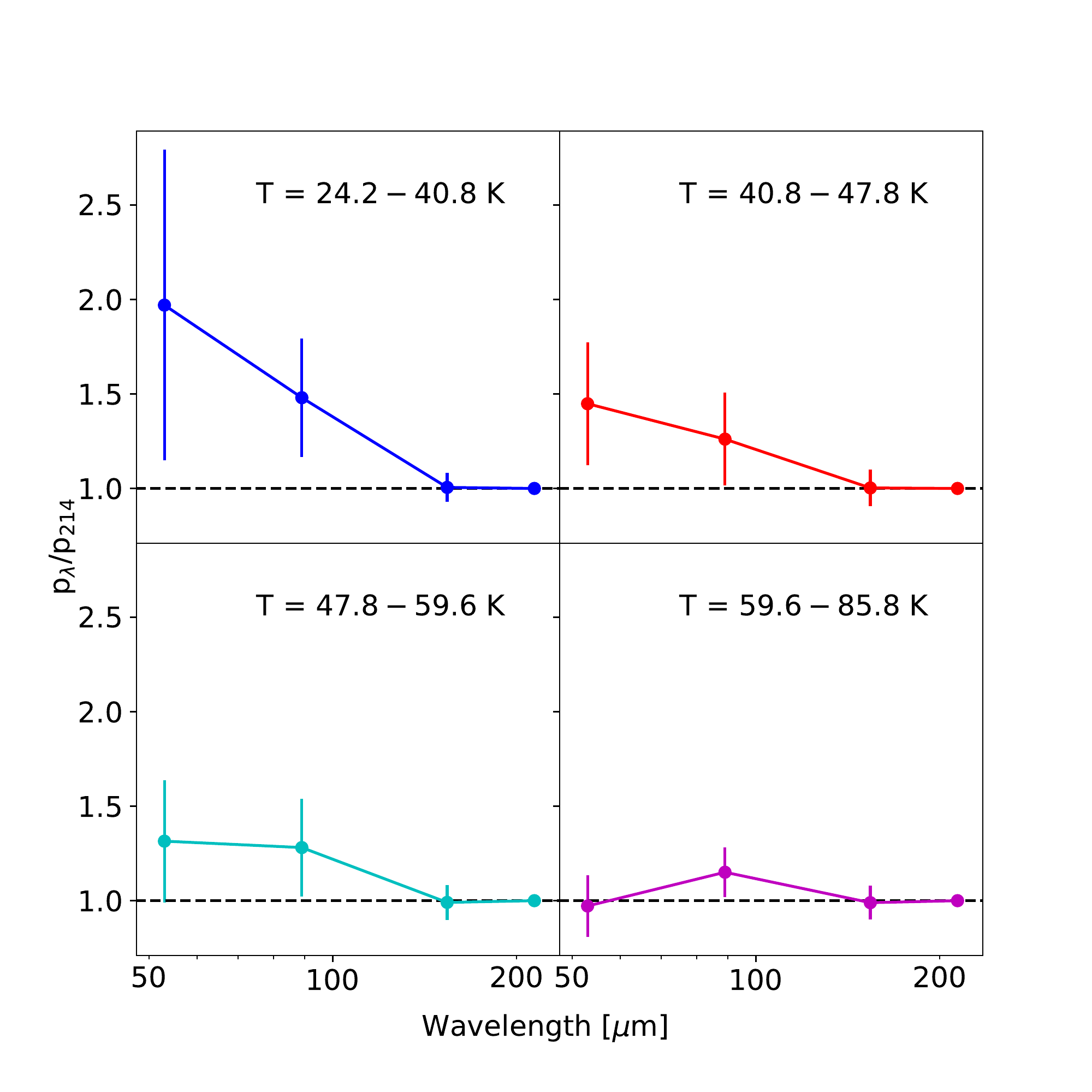}
            \includegraphics[width=3.5in]{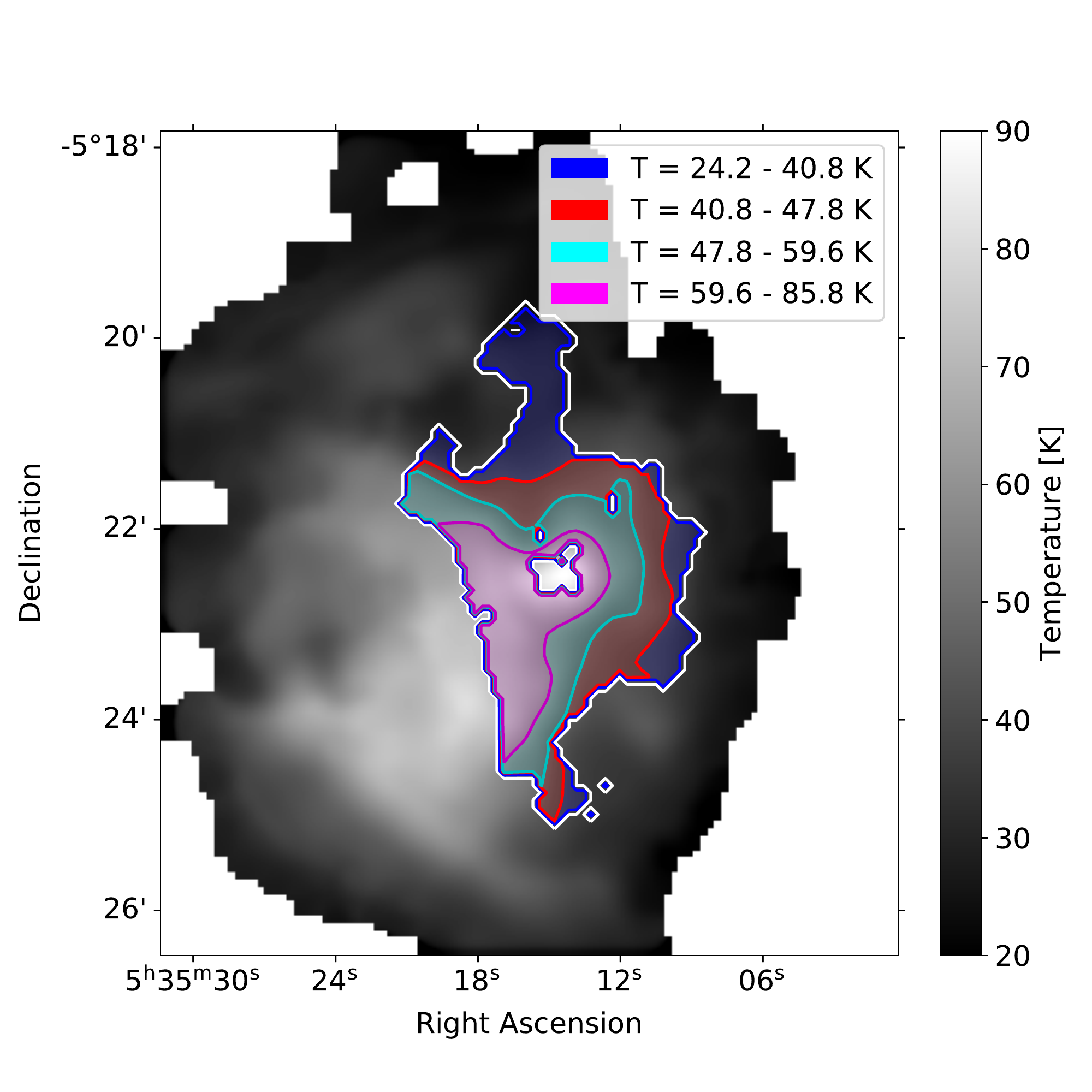}
            \caption{Left: polarization spectra for four equal-pixel areas of the BNKL region. The data were binned into four quartiles based on dust temperature. The ranges of temperature are listed at the top of each plot. Right: spatial maps of the temperature quartiles in the BNKL region. The background is the dust temperature map from \citet{Chuss2019}. The white contour marks the boundary of the BNKL region after polarization cuts.}
            \label{fig:bnkl_temp_quartiles}
        \end{figure}

      Because the BNKL region (see Section \ref{ssec:pspec_region}) is distinguished by high column densities and a very wide range of dust temperatures, we can use it to test our conclusions regarding the positive correlation between $b_l$ and temperature.   We divide the sight lines within the BNKL region into four sets based on temperature, and the polarization spectrum of each such temperature quartile is determined using the methods described in Section \ref{ssec:pspec_region}.  These are shown in Figure \ref{fig:bnkl_temp_quartiles} and demonstrate that the slope of the polarization spectrum is a function of temperature, as suggested earlier.  At low temperatures, the data show clearly falling median spectra; at high temperatures, the spectra are approximately flat. 
       
      We conclude from Figures~\ref{fig:sedparam_linear}--\ref{fig:bnkl_temp_quartiles} and the accompanying discussion that while some of the trends seen in the two-band polarization spectra of $\rho$ Oph A are also seen in our four-band analysis of Orion, our data show a more complex picture. We see a clear temperature trend, but, unlike \citet{Santos2019}, we see no clear trend with column density.

    \subsection{Polarization Ratios across the Cloud}\label{ssec:ratio_maps}
\begin{figure}[htpb]
            \centering
            \includegraphics[trim={0.85cm 8cm 0.8cm 8cm }, clip, width=6in]{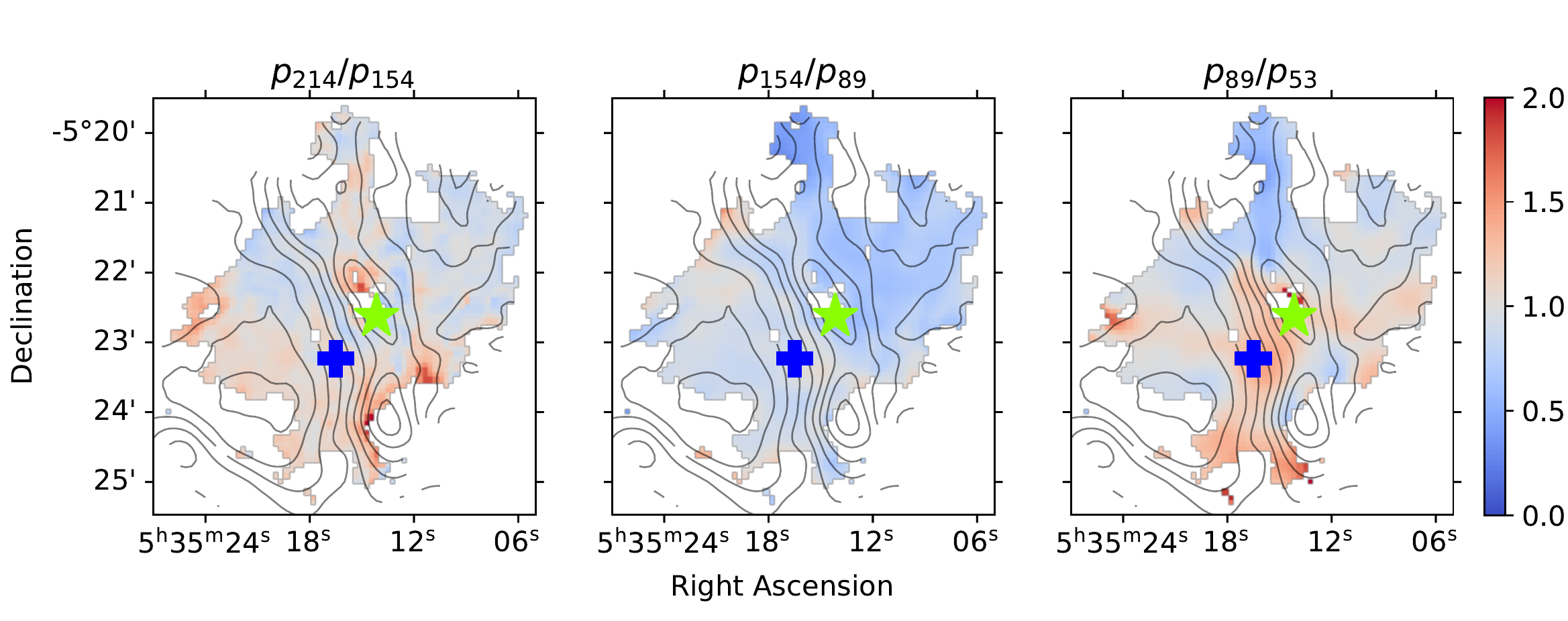}
            \caption{Spatial polarization maps of the OMC-1 cloud: $p_{214}/p_{154}$ (left), $p_{154}/p_{89}$ (center), and $p_{89}/p_{53}$ (right). The green star marks the location of the BN/KL, and the blue cross indicates the center of the Trapezium Cluster. The column density contours are shown for 10 logarithmically spaced intervals in the range $N$(H$_2$) = $10^{21-24}$ cm$^{-2}$.}
            \label{fig:pol_ratio}
        \end{figure}

\begin{figure}[htbp]
            \centering
            \includegraphics[width=6in]{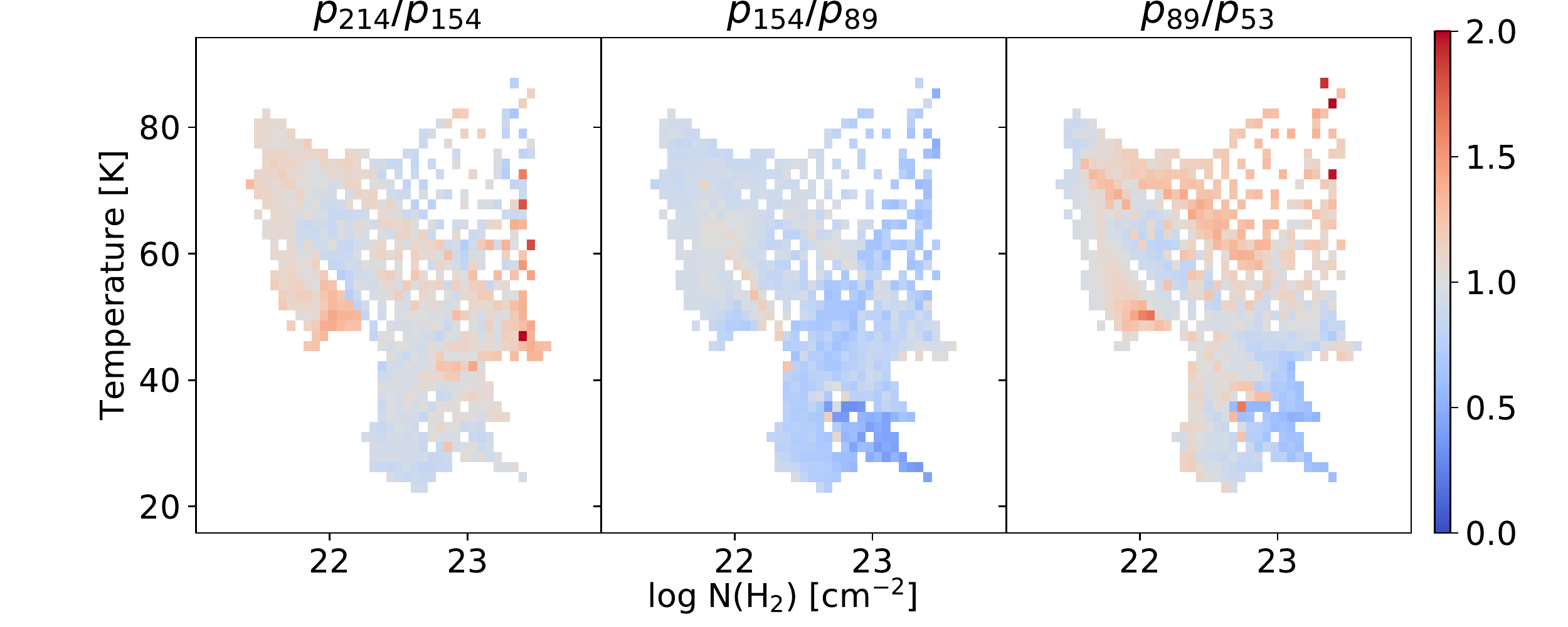}
            \caption{Dependence of neighboring polarization ratios on temperature and column density. The temperature and logarithm of column density are divided into 50 separate bins. The color of each bin represents the median polarization ratio within it.}
            \label{fig:N_T_R}
        \end{figure}

        In this section, we present maps of the ratios of polarization fraction in neighboring wavelength channels. These maps are shown in Figure \ref{fig:pol_ratio}. Note that when calculating the polarization ratio, we use the polarization fraction at the shorter of the two wavelengths as the normalizing value. This shift in notation allows for a direct comparison with \citet{Santos2019}, where values $>1$ ($<1$) indicate rising (falling) spectra. We show these three ratios as functions of temperature and column density in Figure~\ref{fig:N_T_R}. Figures \ref{fig:pol_ratio} and \ref{fig:N_T_R} may be compared, respectively, with the right panel of Figure \ref{fig:fit_params} and the left panel of Figure \ref{fig:N_T_P}. Many of the $b_l$ trends seen in the earlier figures can be noted in these polarization ratio trends, but there are interesting variations across the three polarization ratios. In particular, most of the negative slopes seem to come from the intermediate ratio, $p_{154} / p_{89}$. This can also be seen in Figure \ref{fig:hawc_pspec_overall} and \ref{fig:hawc_pspec_bnkl}.

         \begin{figure}[htbp]
            \centering
            \includegraphics[trim= 0 0.0cm 0.0cm 1.25cm, clip, width=5in]{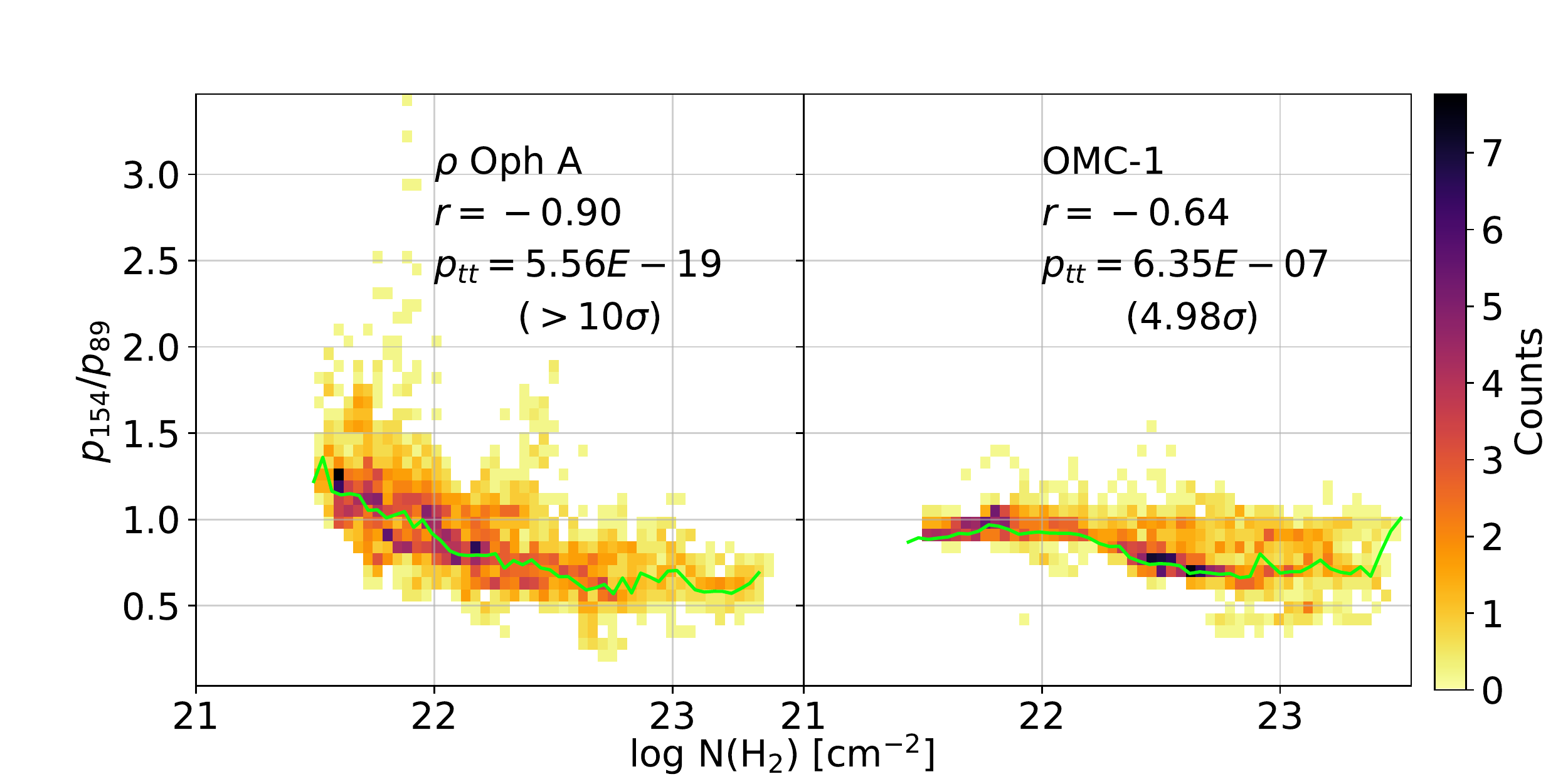}\\
            \includegraphics[trim= 0 0.0cm 0.0cm 1.25cm, clip, width=5in]{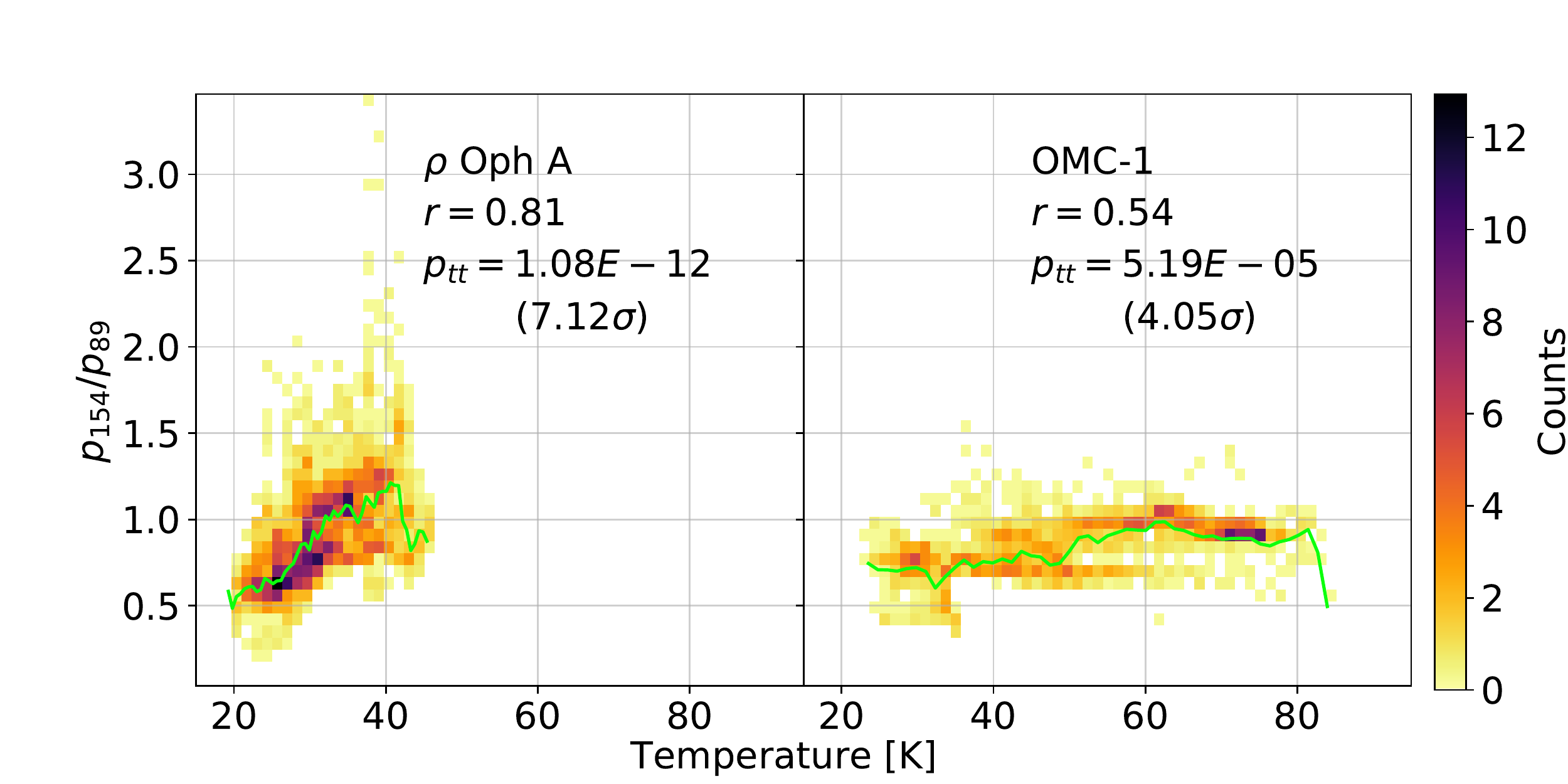}
            \caption{Plots corresponding to Figure 6 (bottom panels) of \citet{Santos2019}. Top: histograms of $p_{154} / p_{89}$ as a function of column density for $\rho$ Oph A (left) and OMC-1 (right). Bottom: histograms of $p_{154} / p_{89}$ as a function of temperature for $\rho$ Oph A (left) and OMC-1 (right). In all plots, the green line corresponds to the median $p_{154} / p_{89}$ ratio for each bin. Pearson correlation coefficients and two-tailed p-values are calculated and listed to quantify the relative strengths of the correlations.}
            \label{fig:santos_fig}
        \end{figure}
        
        We complete this analysis of the OMC-1 region using the prescription described in \citet{Santos2019}, which uses HAWC+ 89 and 154 \micron\ polarimetric observations to probe the polarization spectrum of the $\rho$ Oph A molecular cloud.  To directly compare OMC-1 with their analysis, we follow the format of Figure 6 (bottom panels, histograms of $p_{154} / p_{89}$ binned by column density and temperature) from \citet{Santos2019} for our OMC-1 data and show the results in Figure~\ref{fig:santos_fig}. This figure is similar to Figure \ref{fig:sedparam_linear}, where we compare a property of the polarization spectrum to environmental parameters. However, we only focus on the two wavelengths used in \citet{Santos2019} rather than looking at the average polarization spectrum slope ($b_l$) across all four wavelengths considered above. Figure \ref{fig:santos_fig} shows that in OMC-1 we do see generally larger values of the $p_{154} / p_{89}$ ratio for low column densities and high temperatures, just as \citet{Santos2019} see in $\rho$ Oph A. However, for the case of OMC-1, these trends are not as strong, as evidenced by the Pearson correlation coefficients.
        
\section{Discussion}\label{sec:discussion}

In Sections \ref{ssec:pspec_region} through \ref{ssec:env}, we explored overall trends in our four-band Orion polarization spectra, and we compared these with the corresponding trends seen in $\rho$ Oph A by \citet{Santos2019} using just two wavelengths.  We noted some similarities.  For example, \citet{Santos2019} reports a positive correlation between polarization spectrum slope and temperature, and we find this too, as can be clearly seen in Figure \ref{fig:sedparam_linear} (bottom left) and Figure \ref{fig:bnkl_temp_quartiles}.  Also, Figure \ref{fig:polspec_hawc} shows a clear contrast between the flat spectra of the low-column density TRP region and the generally negatively sloped spectra in the much denser BNKL region.  This is in qualitative agreement with the negative correlation between polarization spectrum slope and column density reported by \citet{Santos2019}.  However, no significant correlation between polarization spectrum slope and column density is seen in Figure \ref{fig:sedparam_linear} (bottom right).  This demonstrates that, along with the similarities between our main results and those of \citet{Santos2019}, there are also some key differences.  In this section, we discuss causes and implications of the similarities and differences between the two data sets.

\citet{Santos2019} developed a simple, spherically symmetric cloud model to explain their two-wavelength FIR polarization spectra for $\rho$ Oph A.  The model relies on values of column density and line-of-sight temperature that they extracted from Herschel data via SED fitting (their fitting method was discussed Section \ref{ssec:env}).  The simple model consists of a spherically symmetric, dense core embedded in a uniform ambient medium of column density $N_b$ and temperature $T_b$. The core has a molecular hydrogen density profile given by a Plummer model with central density $n_0$ and central temperature $T_0$. The temperature in the core is assumed to rise linearly out to radius $R$, from $T_0$ at the center to $T_R$ at the core's edge.  Their intention is to model a core that is heated uniformly from the outside.  In reality, the heating radiation originates from a single high-mass star, Oph S1, and impacts only one side of the core. However, \citet{Santos2019} argue that because they observed primarily the eastern (illuminated) portion of the core, their simple model should provide a reasonable approximation.  

The core model includes a transition radius, $R_t$, inside of which the dust grains are assumed to emit only unpolarized thermal radiation (note that $R_t < R$).  For observations along lines of sight passing within distance $R_t$ of the core center, the polarized intensity originates in the outer layers of the core ($r > R_t$) and in the warm ambient medium surrounding the core, and not from material having $r < R_t$ that emits only unpolarized FIR radiation.  In this way, \citet{Santos2019} are able to incorporate HCE (see Section \ref{sec:intro}) into their model.  The polarization fraction is calculated by integrating a modified blackbody along the line of sight assuming a spatially uniform (but wavelength-dependent) ambient polarization efficiency outside of the transition radius. As shown in their Figure 6, the simple model matches the dependence of the $p_{154} / p_{89}$ ratio on $N_H$ and $T$ quite well.  \citet{Santos2019} conclude that HCE provides a reasonable quantitative explanation for the trends in $p_{154} / p_{89}$ they observed, capturing both the positive correlation of the polarization spectrum slope with temperature and the negative correlation of the polarization spectrum slope with column density.  These correlations can be seen in the two left panels of Figure \ref{fig:santos_fig}, and the corresponding predictions of the model can be seen in Figure~6 of \citet{Santos2019}.  

As stated in Section \ref{sec:intro}, the molecular cloud models of \cite{Bethell2007} are not able to explain negatively sloped FIR spectra.  We speculate that this is due to the lack of strong radiation sources, as these authors included no radiation other than the standard interstellar radiation field, assumed to originate from field stars surrounding their model cloud.  Besides \citet{Santos2019} and \citet{Bethell2007}, no molecular cloud models including HCE have produced polarization spectra for comparison with data such as ours.
        
The OMC-1 region has dense cores and high-mass stars, just as $\rho$ Oph A does, so the above-mentioned similarities between our polarization spectrum results and those of \citet{Santos2019} suggest that HCE is operating in Orion just as it operates in $\rho$ Oph A.  But what explains the differences?  In particular, why would Orion not show unambiguously negative polarization spectrum slopes for the highest-column-density sight lines, as predicted by the model of \citet{Santos2019} and as seen in $\rho$ Oph A? The generally higher level of high-mass star formation activity in Orion as compared to $\rho$ Oph A may drive grain alignment even for regions of very high column density, thus explaining why in Orion, but not in $\rho$ Oph A, positively sloped spectra are commonly seen for $N$(H$_2$) $> 10^{23}$ cm$^{-2}$ (compare the upper left panel of Figure \ref{fig:santos_fig} with the bottom right panel in Figure \ref{fig:sedparam_linear}). OMC-1 has a prevalence of high-luminosity embedded sources, and RATs from these stars can maintain a high degree of alignment even deep in the cloud where the column density is high.  This is consistent with one of the principal conclusions of \citet{Chuss2019}, in which it was found that loss of grain alignment at high column densities was not required to explain the anticorrelation of polarization fraction with intensity (used as a rough proxy for column density).

This evidence leads to the conclusion that it is the radiative environment, as traced by the dust temperature -- not the column density -- that appears to determine the alignment of grains. Put another way, the loss of grain alignment efficiency that makes HCE possible is attributable to the absence of radiation (due to shielding) rather than to any direct effect of increasing gas density.  Two such direct effects that were discussed in Section \ref{sec:intro} are disruption of grain alignment by gas particle collisions and changes in grain shape -- tending toward rounder grains -- due to dust grain coagulation in dense regions.  It is of course possible that radiation may have a direct effect on grain shape, such as via its effect on ice mantles.  Our point is not that grain shape effects cannot be causing the HCE but rather that polarization efficiency effects purely driven by gas density are disfavored by our observation that high-density sight lines often appear to lack HCE. 

As discussed in Section \ref{ssec:ratio_maps} and displayed in Figure \ref{fig:santos_fig}, we restricted the analysis to just the two intermediate wave bands of our four-band OMC-1 polarization spectra in order to make a more direct comparison between OMC-1 and $\rho$ Oph A.  This showed some consistency between the two clouds; however, the two-band ratio analysis on its own is insufficient to reach the conclusions above.  This motivates the need for future observations to examine more clouds with the full range of HAWC+ bands. For comparison with these observations, we require models for the HCE that make predictions for the full range of bands.  In addition, as a complementary technique, we suggest that the $p$ versus $I$ relationship should be explored in $\rho$ Oph A (see discussion earlier in this section) to better illuminate the similarities and differences between $\rho$ Oph A and OMC-1. These and other observations that probe the extent to which loss of polarization efficiency occurs in star forming clouds and the extent to which this is correlated with physical parameters like temperature and $N$(H$_2$) should lead to improved methods for estimating magnetic field strength in these clouds (see Section \ref{sec:intro}).  This in turn may lead to a better overall understanding of the physical processes involved in star and planet formation.     

\section{Summary}\label{sec:summary}
The prevailing explanation for a falling FIR polarization spectrum is a superposition of multiple temperatures along the line of sight where cooler grains having low polarization efficiency reside in denser regions and the warmer grains with high polarization efficiency reside in less dense regions that are more exposed to radiation from field stars or young stellar objects \citep{Hildebrand1999}. We refer to this superposition effect as the ``HCE''.

We have used continuum HAWC+/SOFIA polarization maps at 53, 89, 154, and 214 \micron~to study polarization spectra in the OMC-1 star-forming region.  The large number of independent sight lines in OMC-1 allowed us to study spatial variations in polarization spectra across this heterogeneous cloud using a variety of complementary techniques.  Our principal results are as follows:
    \begin{enumerate}
        \item We find evidence of a flat spectrum within the TRP region to within about 5\%.
        \item We find that the polarization spectrum within the BNKL region is highly variable.  In the cooler regions, we observe falling spectra; in the warmer regions, the spectra tend to be flatter. 
        \item We explore how the slope of the polarization spectrum depends on both column density and line-of-sight temperature.  Polarization spectrum slopes were found by fitting linear forms to each sight lines' individual polarization measurements. We find a clear positive correlation between polarization spectrum slope and temperature but no significant correlation between spectral slope and column density.
        \item Our explanation for the trends we find is consistent with that of \citet{Santos2019}, namely, the disappearance of HCE for sight lines lacking cold, poorly aligned grains. 
        \item Our analysis indicates that HCE is more likely explained by RAT theory as opposed to changes in dust grain shape or alignment efficiency driven purely by density. This conclusion should be further tested by measuring the polarization spectra of additional clouds.  In doing so, all available wavelengths should be measured to obtain a complete understanding of the variation of the shape of the spectrum as a function of physical parameters. 
    \end{enumerate}

\acknowledgments
\section*{Acknowledgments}
This work is based on observations made with the NASA/DLR Stratospheric Observatory for Infrared Astronomy (SOFIA). SOFIA is jointly operated by the Universities Space Research Association, Inc. (USRA), under NASA contract NAS2-97001, and the Deutsches SOFIA Institut (DSI) under DLR contract 50 OK 0901 to the University of Stuttgart. Financial support for this work was provided by NASA through awards SOF 05-0038 and SOF 05-0018 issued by USRA to Villanova University and awards SOF 06-0116 and SOF 07-0147 issued by USRA to Northwestern University. Portions of this work were carried out at the Jet Propulsion Laboratory, operated by the California Institute of Technology under a contract with NASA. Parts of the analysis were performed using the Clusty Computing Facility in the Villanova University Department of Astrophysics and Planetary Science. We thank Andrej Pr\v{s}a for his support in leading and maintaining this resource. The authors would like to thank the anonymous referee for their very detailed and helpful comments.

\software{ \texttt{python, Ipython} \citep{Perez2007}, \texttt{numpy} \citep{vanderWalt2011}, \texttt{scipy} \citep{2020SciPy}, \texttt{matplotlib} \citep{Hunter2007}, \texttt{astropy} \citep{astropy:2013, astropy:2018}} 

\facility{SOFIA (HAWC+)}
\bibliographystyle{aasjournal}
\bibliography{OMC1}

\appendix
\section*{Comparison of HAWC+ Polarization across All Bands}
\begin{figure}[htbp]
            \centering
            \includegraphics[trim={2.5cm 3cm 0.5cm 4.cm }, clip, width=6in]{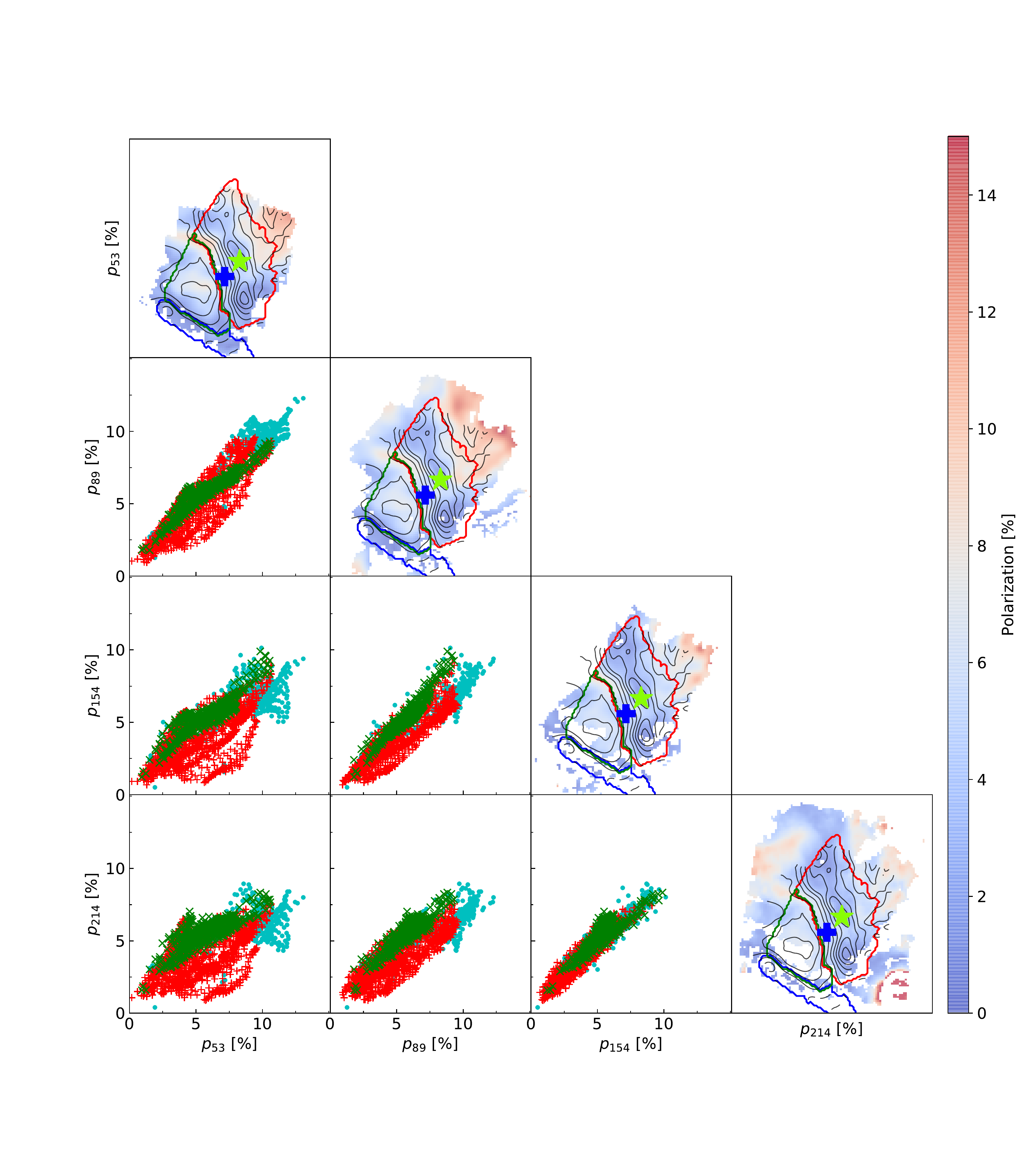}
            \caption{Corner plot of all HAWC+ polarization ratios. The polarization percent maps are shown on the diagonal for each filter with the three region masks overlaid. The marker colors correspond to pixels within each of the regions; cyan color markers correspond to pixels that do not lie inside of these regions. The points in the polarization ratio plots are for sight lines where the range in polarization angle is less than 15$^\circ$.}
            \label{fig:hawc_corner}
        \end{figure}

We present a corner-plot representation of the HAWC+ polarization ratios for each combination of wavelengths. The data are colored by region (see Section \ref{sec:data}) to show any difference in polarization efficiencies over the cloud. This is shown in Figure \ref{fig:hawc_corner}.
    
\end{document}